\documentclass[english,american,journal,twocolumn]{IEEEtran}
\usepackage[T1]{fontenc}
\usepackage[latin9]{inputenc}
\usepackage{rotating}
\usepackage{amsmath}
\usepackage{graphicx}

\makeatletter

\providecommand{\tabularnewline}{\\}

 \let\oldforeign@language\foreign@language
 \DeclareRobustCommand{\foreign@language}[1]{%
   \lowercase{\oldforeign@language{#1}}}

\usepackage{fancyhdr}

\usepackage{pstricks, pst-node, pst-plot, pst-circ}
\usepackage{eurosym}
\usepackage{color}
\usepackage{graphicx}

\usepackage{psfrag}

\usepackage{xcolor}
\usepackage{xfrac}

\usepackage{url}

\usepackage{amsmath}
\usepackage{amssymb}
\usepackage{array}

\usepackage{booktabs}

\usepackage{multirow}

\usepackage{times,epsfig}
\usepackage{epstopdf}
\usepackage{rotating}

\usepackage{textcomp}

\usepackage{pgfplots}
\pgfplotsset{compat=newest}
\pgfplotsset{plot coordinates/math parser=false}

\newcommand{\scalable}{scalable}
\newcommand{\downscale}{down-scale}
\newcommand{\scalability}{scalability}

\newcommand{\wt}{wavelet transform}
\newcommand{\wl}{wavelet lifting}
\newcommand{\ls}{lifting structure}
\newcommand{\LS}{Lifting structure}

\newcommand{\PStep}{prediction step}
\newcommand{\UStep}{update step}

\newcommand{\subsampling}{sub-sampling}

\newcommand{\Lpass}{lowpass}
\newcommand{\Lband}{\Lpass{} band}
\newcommand{\Lpartband}{\Lpass{} sub-band}
\newcommand{\Lpartbands}{\Lpass{} sub-bands}
\newcommand{\Lbandcoeff}{\Lpass{} coefficient}
\newcommand{\Hpass}{highpass}
\newcommand{\Hpartband}{\Hpass{} sub-band}
\newcommand{\Hbandcoeff}{\Hpass{} coefficient}

\newcommand{\sigf}{f}
\newcommand{\HP}{\text{H\hspace{-0.25mm}P}}
\newcommand{\LP}{\text{L\hspace{-0.25mm}P}}

\newcommand{\blockbased}{block-based}
\newcommand{\Meshbased}{Mesh-based}
\newcommand{\meshbased}{mesh-based}

\newcommand{\thresDeterJacobian}{\Delta}
\newcommand{\meshsmoothness}{S}
\newcommand{\meshsr}{d_\text{sr}}
\newcommand{\meshbs}{s}
\newcommand{\regterm}{r}
\newcommand{\reglambda}{\lambda}
\newcommand{\meshmetric}{J}
\newcommand{\meshmetricold}{\meshmetric_{\text{soa}}}
\newcommand{\meshmetricprop}{\meshmetric_{\text{prop}}}

\newcommand{\fig}[1]{Fig.~#1}
\newcommand{\Fig}[1]{Fig.~#1}
\newcommand{\tab}[1]{Tab.~#1}
\newcommand{\Tab}[1]{Tab.~#1}
\newcommand{\sect}[1]{Section~#1}

\newcommand{\jptwok}{JPEG~2000}
\newcommand{\specktwod}{SPECK}
\newcommand{\Legall}{LeGall~5/3 wavelet}

  \newlength\figureheight
    \newlength\figurewidth
\makeatother

\usepackage{babel}
\begin{document}

\title{Temporal Scalability of Dynamic Volume Data\\
using Mesh Compensated Wavelet Lifting}

\author{Wolfgang~Schnurrer, Niklas~Pallast, Thomas~Richter, and André~Kaup,~\IEEEmembership{Fellow,~IEEE}%
\thanks{The authors are with the Chair of Multimedia Communications and Signal Processing, Friedrich-Alexander-University Erlangen-Nürnberg (FAU), Cauerstr.~7, 91058~Erlangen, Germany (e-mail: wolfgang.schnurrer@fau.de; niklas.pallast@fau.de; thomas.richter@fau.de; andre.kaup@fau.de).%
}}

\maketitle

\markboth{~}{Schnurrer \MakeLowercase{\textit{et al.}}: Temporal Scalability
of Dynamic Volume Data using Mesh Compensated Wavelet Lifting}
\begin{abstract}
Due to their high resolution, dynamic medical 2D+t and 3D+t volumes
from computed tomography (CT) and magnetic resonance tomography (MR)
reach a size which makes them very unhandy for teleradiologic applications.
 A \scalable{} representation offers the advantage of a \downscale{}d
version which can be used for orientation or previewing, while the
remaining information for reconstructing the full resolution is transmitted
on demand.  The \wt{} offers the desired \scalability{} but a very
high quality of the \Lband{} is crucial in order to use it as a \downscale{}d
representant.  

We propose an approach based on compensated \wl{} for obtaining a
\scalable{} representation of dynamic CT and MR volumes with very
high quality. The mesh compensation is feasible to model the displacement
in dynamic volumes which is mainly given by expansion and contraction
of tissue over time. To achieve this, we propose an optimized estimation
of the mesh compensation parameters to optimally fit for dynamic volumes.
Within the \ls{}, the inversion of the motion compensation is crucial
in the update step. We aim to take this inversion directly into account
during the estimation step and can improve the quality of the \Lpartband{}
by 0.63~dB and 0.43~dB on average for our tested dynamic CT and
MR volumes at the cost of a increase of the rate by 2.4\% and 1.2\%
on average. \end{abstract}

\begin{IEEEkeywords}
Scalability, Discrete Wavelet Transforms, Motion Compensation, Computed
Tomography, Magnetic Resonance Tomography, Signal Analysis
\end{IEEEkeywords}

\IEEEpeerreviewmaketitle{}

\section{Introduction}

A \scalable{} representation of huge volume data offers the advantage
of providing a \downscale{}d version much faster. This \downscale{}d
representation can be utilized for faster browsing or orientation
within the volume. Particularly when the volumes have to be transmitted,
e.g., when the acquisition is done at a different location than the
diagnosis or the volume data needs to be accessed by doctors, e.g.,
at rural areas. In the medical environment, lossless reconstruction
is a crucial condition, although lossy compression has already been
analyzed \cite{loose2008}. 

\begin{figure}
\begin{center}
\psfragscanon
\psfrag{ti}{{\tiny $t$}}
\psfrag{x}{{\tiny $x$}}
\psfrag{y}{{\tiny $y$}}
\psfrag{t}{temporal \scalable{} representation}
\psfrag{s}{high quality \Lpartband{}}
\psfrag{e}{encode}
\psfrag{h}{$\HP$}
\psfrag{l}{$\LP$}
\psfrag{hp}{$\HP$}
\psfrag{lp}{$\LP$}
\psfrag{bs}{bitstream}
\psfrag{m}{$\begin{array}{c}\text{\meshbased}\\[-1ex]\text{compensated}\\[-1ex]\text{\wl{}}\end{array}$}

\includegraphics[height=3cm]{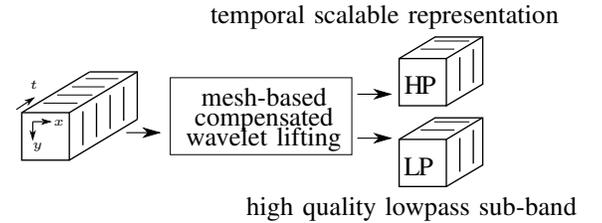}

\psfragscanoff
\end{center}

\protect\caption{\label{fig:block-diagram-overview}The block diagram shows the considered
scenario. Compensated \wl{} is utilized to obtain a temporal \scalable{}
representation of a dynamic volume. \Meshbased{} methods are feasible
to compensate the displacement in dynamic medical volumes and thus
increase the quality of the \Lpartband{} to serve as \downscale{}d
version.}
\end{figure}

A wavelet-based approach as shown in \fig{\ref{fig:block-diagram-overview}}
can provide a \scalable{} representation as well as the necessary
prefect reconstruction. In teleradiology, this enables new scenarios
where a \downscale{}d representation can be transmitted faster. If
only a specific part with a higher resolution is needed, solely the
corresponding additional information has to be transmitted. By combining
a high quality \scalable{} representation with anatomic information
\cite{cavallaro2011region} the specific part can be provided lossless
while the remaining part of the volume is available in a coarser representation.
This volume of interest coding can be obtained by optimized coefficient
coding \cite{sanchez2010}. Dependent on the application, it is additionally
possible to reconstruct the original volume completely~\cite{calderbank1997}.

A wavelet-based approach offers a fundamentally different temporal
scalability compared to a hybrid coding approach like  H.265/HEVC~\cite{hevc_overview},
the current state of the art for coding video data. Using the hybrid
approach, temporal \scalability{} can be obtained by choosing an
appropriate prediction scheme~\cite{sullivan2013shvc} supporting
the skip of, e.g., every other frame. This corresponds to a \subsampling{}
without prior \Lpass{} filtering. In contrast to that the \Lpartband{}
from a \wt{} results from \subsampling{} after lowpass filtering.
In medical applications, this comes with the advantage that structures,
only visible within one original frame, do not disappear in the \scalable{}
representation due to \subsampling{}. Displacements within the original
volume that cannot be modeled by the compensation method are still
contained within the \Lpartband{}.

In the medical environment, the Digital Imaging and Communications
in Medicine (DICOM) standard \cite{dicom2012} is used for communication.
This standard contains several coding methods for image data. In practice,
usually a frame by frame coding using lossless JPEG~\cite{wallace1992}
is used for coding.

The \Lpartband{} of a \wt{} is computed as weighted average. By
design, displacement in the signal can cause a blurry \Lpartband{}
with ghosting artifacts. This drawback can be addressed by incorporating
feasible compensation methods directly into the transform. In video
coding, a compensated \wt{} in temporal direction is called Motion
Compensated Temporal Filtering (MCTF) \cite{Ohm1994}. Video sequences
mostly contain translatory motion. Consequently, \blockbased methods
are mainly used for compensation~\cite{Ohm1994,Choi1999}. However,
the motion model of translatory movement of objects does not fit well
for dynamic medical volumes~\cite{nosratinia1996}. Within \wl{}
of video sequences, \meshbased{} compensation has also proven its
better suitability~\cite{secker2003}. The displacement over time
is mainly given by expansion and contraction of tissue. \Meshbased{}
compensation methods have shown their ability to model the deforming
displacement well \cite{schnurrer2012mmsp,schnurrer2014}. \Meshbased{}
methods have also shown to be well suited to model the deformation
within dynamic medical volumes in other applications, e.g., for image
registration \cite{Frakes2008}.  

The inversion of the \meshbased{} compensation is, however, most
often not needed and thus not considered when estimating the parameters
of the mesh deformation. However, within the update step of wavelet-lifting,
the compensation has to be inverted \cite{bozinovic2005}. This inversion
is not only necessary but very important in order to move the remaining
high frequent information to the corresponding position for the \Lpass{}
frame. Consequently, it is also very important for the quality of
the \Lpartband{}. So far in the current literature, only the compensation
is considered during the estimation of the parameters for the \meshbased{}
compensation.

The characteristic of the dynamic medical volumes and their containing
displacement can be used directly for providing properties of the
mesh. The deforming displacement to be compensated results from expansion
and contraction of tissue over time. The displacement can be described
as smooth without discontinuities, i.e., there are no discontinuities
in the mesh either. A regularizer can be used to prefer similar movement
of neighboring grid points and thus a smooth motion vector field~\cite{heising2001}.
The regularizer also prevents small local movements in the mesh due
to noise. Medical images usually cover the complete displacement,
so the border grid points of the mesh can only move along the border
but remain there.

In this article, we propose an approach to obtain a \scalable{} representation
of dynamic volumes based on mesh compensated \wl{}. The approach
is based on our work previously published: In~\cite{schnurrer2012vcip},
a compensated \wt{} in $z$-direction is analyzed. In~\cite{schnurrer2012mmsp},
we compared triangular and quadrilateral meshes where quadrilateral
meshes show a better performance for a compensated \wt{} in temporal
direction of dynamic CT volumes. In~\cite{schnurrer2014}, 3-D mesh
compensation is examined.

In this article, we introduce a novel criterion for the \meshbased{}
motion estimation and a new method to evaluate the smoothness of the
mesh deformation. The article is outlined as follows. Section \ref{sec:Compensated-Wavelet-Lifting}
starts with a brief review of compensated \wl{} and introduces \meshbased{}
compensation in \ref{sec:Mesh-based-compensation}. The estimation
for the mesh deformation follows in section \ref{sec:Proposed-Grid-Point-ME}
with our proposed adaptions and our proposed metric. Simulation results
are presented in section \ref{sec:Simulation-Results}. Section \ref{sec:Conclusion}
gives the conclusions.

\section{\label{sec:Compensated-Wavelet-Lifting}Mesh Compensated Wavelet
Lifting}

The \ls{} is an efficient way to compute a \wt{}. Therefore, wavelet
filters are factorized into the \ls{} \cite{daubechies1998}. The
\ls{} consists of two steps. In the prediction step, the \Hpartband{}
is computed according to
\begin{equation}
\HP_{t}=\sigf_{2t}-\sigf_{2t-1}\label{eq:H-Haar-lifting}
\end{equation}
where $\sigf_{t}$ denotes a frame at time step $t$ of the signal
to be transformed.

In the update step, the \Lpartband{} $\LP_{t}$ is computed according
to 
\begin{equation}
\LP_{t}=\sigf_{2t-1}+\left\lfloor \frac{1}{2}\HP_{t}\right\rfloor \label{eq:L-Haar-lifting}
\end{equation}
utilizing the already computed \Hpartband{} coefficients $\HP_{t}$
in (\ref{eq:H-Haar-lifting}).

In (\ref{eq:L-Haar-lifting}), the relationship between \Hpartband{}
and \Lpartband{} becomes immediately visible. In addition to that,
the \ls{} offers several more advantages. An integer transform can
be obtained very easily by introducing rounding operators \cite{calderbank1997}
around the fractional parts as in (\ref{eq:L-Haar-lifting}). Thereby,
rounding errors are avoided. The perfect reconstruction property makes
this transform feasible and very interesting for medical applications
as it does not only offer a \scalable{} representation but also the
possibility to reconstruct the original signal without loss.. Furthermore,
arbitrary compensation methods can be implemented directly into the
transform \cite{garbasTCSVT} without loosing the prefect reconstruction
property. Blurriness and artifacts within the \Lpartband{} caused
by displacement over time can be avoided by utilizing a feasible compensation
method.

\begin{figure}
\psfragscanon
\psfrag{r}{$\sigf_{2t-1}$}
\psfrag{c}{$\sigf_{2t}$}
\psfrag{h}{$\HP_t$}
\psfrag{l}{$\LP_t$}
\psfrag{p}{$p_{2t}$}
\psfrag{u}{$u_{2t}$}
\psfrag{m}{MC}
\psfrag{i}{IMC}
\psfrag{t}{time $t$}
\psfrag{s}{$-$}
\psfrag{a}{$+\frac{1}{2}$}

\includegraphics[width=1\columnwidth]{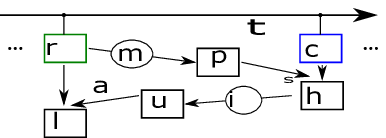}

\psfragscanoff

\protect\caption{\label{fig:compensated-wavelet-lifting}\LS{} for the Haar \wt{}
with compensation (MC). The sequence $\sigf$ is decomposed into a
\Hpartband{} and a \Lpartband{}. In the update step of the \ls{},
the compensation has to be inverted (IMC).}

\end{figure}

\subsection{Compensated Wavelet Lifting}

For obtaining temporal \scalability{}, the \wt{} is applied in temporal
direction. The \Hbandcoeff{}s $\HP_{t}$ of a compensated \wt{}
are computed in the \PStep{} according to
\begin{equation}
\HP_{t}=\sigf_{2t}-\left\lfloor \mathcal{W}_{2t-1\rightarrow2t}\left(\sigf_{2t-1}\right)\right\rfloor .\label{eq:H-Haar}
\end{equation}
Instead of the original frame $\sigf_{2t-1}$, a predictor is subtracted
from $\sigf_{2t}$, denoted by the warping operator $\mathcal{W}_{2t-1\rightarrow2t}$
\cite{garbasTCSVT}. The compensation is called MC in \fig{\ref{fig:compensated-wavelet-lifting}}.
The result of the warping operator $\mathcal{W}_{2t-1\rightarrow2t}$
is a predictor $p_{2t}$ for the current frame $\sigf_{2t}$ based
on the reference frame $\sigf_{2t-1}$ and can be written as
\begin{eqnarray}
p_{2t}\left(x_{c},y_{c}\right) & = & \mathcal{W}_{2t-1\rightarrow2t}\left(f_{2t-1}\right)\label{eq:warping}\\
 & = & \tilde{\sigf}_{2t-1}\left(x_{r}\left(x_{c},y_{c}\right),y_{r}\left(x_{c},y_{c}\right)\right)\nonumber 
\end{eqnarray}
 where $x_{r}\left(x_{c},y_{c}\right)$ and $y_{r}\left(x_{c},y_{c}\right)$
contain the mapping for pixels within the current frame to the pixels
in the reference frame. For a given position in the current frame,
$x_{r}\left(x_{c},y_{c}\right)$ and $y_{r}\left(x_{c},y_{c}\right)$
provide the corresponding position in the reference frame. The tilde
denotes to the necessary interpolation to obtain intensity values
at non-integer positions.

In the \UStep{} of the compensated \wt{}, the \Lbandcoeff{}s $\LP_{\text{Haar},t}$
are computed by 
\begin{equation}
\LP_{t}=\sigf_{2t-1}+\left\lfloor \frac{1}{2}\mathcal{W}_{2t\rightarrow2t-1}\left(\HP_{t}\right)\right\rfloor .\label{eq:L-Haar}
\end{equation}
The index of $\mathcal{W}$$ $ in (\ref{eq:L-Haar}) shows the inverse
compensation in the update step. This is crucial to achieve an equivalent
\wt{}~\cite{bozinovic2005}. The inversion of the compensation warps
the remaining high frequency structures within $\HP_{t}$ to their
corresponding positions within $\sigf_{2t-1}$ to provide a reasonable
update. The inversion is denoted by IMC in \fig{\ref{fig:compensated-wavelet-lifting}}.
For a compensated \wt{}, the \meshbased{} compensation method has
the advantage that it is invertible \cite{secker2003,heising2001}.

\subsection{\label{sec:Mesh-based-compensation}\Meshbased{} Compensation}

\begin{figure}
\hfill{}\includegraphics[height=4cm]{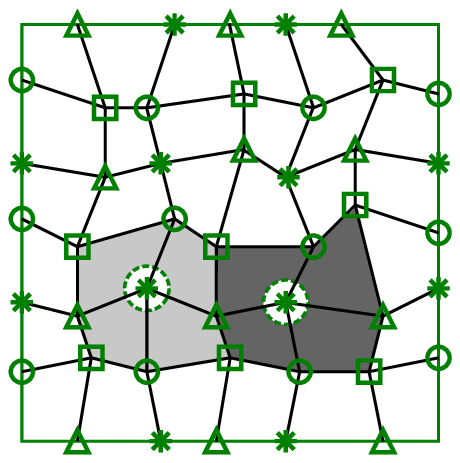}\hfill{}\includegraphics[height=4cm]{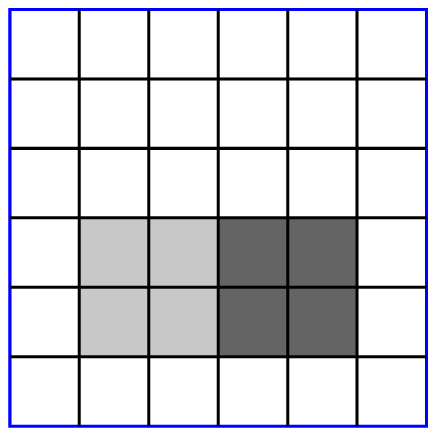}\hfill{}

\hfill{}reference\hfill{}current\hfill{}

\protect\caption{\label{fig:mesh-based-compensation-overview}The green deformed mesh
on the left corresponds to the reference frame $\sigf_{2t-1}$ and
the blue mesh to the current frame $\sigf_{2t}$. The prediction $p_{2t}$
is computed by applying the bilinear transform for each quadrilateral
$i$ in $\sigf_{2t-1}$ according to it's deformation. The update
$u_{2t}$ is obtained by applying the inverse bilinear transform for
every quadrilateral in $\sigf_{2t}$ respectively.}
\end{figure}

In this Section, the \meshbased{} compensation and its inversion
is explained in assumption of the mesh parameters, i.e., the motion
vectors of the grid points, to be known. The estimation of the grid
point motion follows in \sect{\ref{sec:Proposed-Grid-Point-ME}}.
In general, a \meshbased{} compensation is computed according to
the deformation of a mesh. Therefore, one mesh is laid over the reference
frame and one is laid over the current frame, see \fig{\ref{fig:mesh-based-compensation-overview}}.
The mesh of the reference frame is deformed as shown in \fig{\ref{fig:mesh-based-compensation-overview}}
on the left. The predictor for the current compensation is computed
by warping each deformed quadrilateral in the reference frame to its
corresponding quadrilateral in the current frame, exemplarily shown
for the gray quadrilaterals. The connectivity of the mesh represents
the image transform used. A triangular mesh leads to an affine transform
\cite{nakaya1994} while a quadrilateral mesh leads to a bilinear
transform~\cite{sullivan1991}. The smoother motion vector field
of a quadrilateral mesh showed to be more feasible for the compensation
of the deformation within dynamic medical volumes~\cite{schnurrer2012mmsp}.

Several properties for the \meshbased{} compensation can be derived
from the assumptions for the displacement over time. The displacement
results from expansion and contraction of tissue over time. This deformation
corresponds to a smooth displacement without motion discontinuities.
There will be no change in the mesh topology like breaking apart at
a motion boundary. Displacement occurs only within a frame, so the
outer shape of the mesh remains on the frame border. This means the
four grid points at the frame corner remain fixed. Grid points on
the left and right border may move up and down only while grid points
on the upper and lower border may move left and right only. Thereby
unconnected pixels \cite{Ohm1994} do not occur.

The \meshbased{} compensation is a feasible approach and can also
be interpreted as a \subsampling{} of a dense smooth motion vector
field. The grid points represent the sampling points and the motion
vectors in between are obtained by bilinear interpolation.

\begin{figure}
\begin{center}
\psfragscanon

\psfrag{x}{$x$}
\psfrag{y}{$y$}

\psfrag{c11}{$\left(\hspace{-2mm}\begin{array}{c}
x_{c,11}\\[-1ex]
y_{c,11}
\end{array}\hspace{-2mm}\right)$}

\psfrag{u}{$u$}
\psfrag{v}{$v$}

\psfrag{uc11}{$\left(\hspace{-2mm}\begin{array}{c}
u_{c,11}\\[-1ex]
v_{c,11}
\end{array}\hspace{-2mm}\right)\hspace{-1.6mm}=\hspace{-1.6mm}\left(\hspace{-2mm}\begin{array}{c}
0\\[-1ex]
0
\end{array}\hspace{-2mm}\right)$}

\psfrag{uc12}{$\left(\hspace{-2mm}\begin{array}{c}
u_{c,12}\\[-1ex]
v_{c,12}
\end{array}\hspace{-2mm}\right)\hspace{-1.6mm}=\hspace{-1.6mm}\left(\hspace{-2mm}\begin{array}{c}
n_u\\[-1ex]
0
\end{array}\hspace{-2mm}\right)$}

\psfrag{uc21}{$\left(\hspace{-2mm}\begin{array}{c}
u_{c,21}\\[-1ex]
v_{c,21}
\end{array}\hspace{-2mm}\right)\hspace{-1.6mm}=\hspace{-1.6mm}\left(\hspace{-2mm}\begin{array}{c}
0\\[-1ex]
n_v
\end{array}\hspace{-2mm}\right)$}

\psfrag{uc22}{$\left(\hspace{-2mm}\begin{array}{c}
u_{c,22}\\[-1ex]
v_{c,22}
\end{array}\hspace{-2mm}\right)\hspace{-1.6mm}=\hspace{-1.6mm}\left(\hspace{-2mm}\begin{array}{c}
n_u\\[-1ex]
n_v
\end{array}\hspace{-2mm}\right)$}

\psfrag{ur11}{$\left(\hspace{-2mm}\begin{array}{c}
u_{r,11}\\[-1ex]
v_{r,11}
\end{array}\hspace{-2mm}\right)$}

\psfrag{ur12}{$\left(\hspace{-2mm}\begin{array}{c}
u_{r,12}\\[-1ex]
v_{r,12}
\end{array}\hspace{-2mm}\right)$}

\psfrag{ur21}{$\left(\hspace{-2mm}\begin{array}{c}
u_{r,21}\\[-1ex]
v_{r,21}
\end{array}\hspace{-2mm}\right)$}

\psfrag{ur22}{$\left(\hspace{-2mm}\begin{array}{c}
u_{r,22}\\[-1ex]
v_{r,22}
\end{array}\hspace{-2mm}\right)$}

\includegraphics[height=4cm]{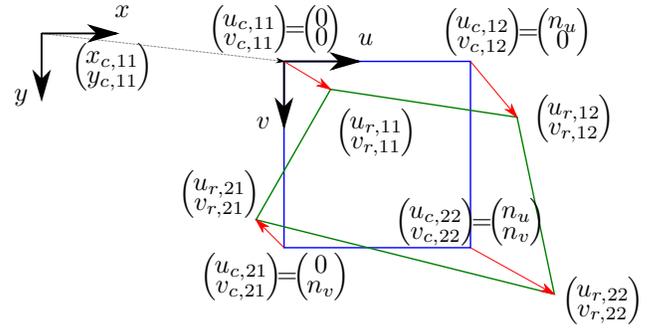}

\psfragscanoff
\end{center}

\protect\caption{\label{fig:bilinear-transform-patch}Diagram of the bilinear transform
for a quadrilateral showing the local coordinate system. The local
coordinate system has its origin in the upper left of the blue quadrilateral,
i.e., within the current frame. The coordinates of the quadrilateral
within the current frame are labeled by $\left(u_{c},v_{c}\right)$
while the coordinates of the green deformed quadrilateral within the
reference frame are labeled by $\left(u_{r},v_{r}\right)$. The red
arrows show the motion vectors of the grid points and thus describe
the deformation of the mesh.}
\end{figure}

For each quadrilateral, the deformation is given by the image transform
that corresponds to the grid point movement. Considering the $i$-th
quadrilateral, the coordinates in the reference frame $\left(x_{r},y_{r}\right)$
can be computed from the coordinates in the current frame $\left(x_{c},y_{c}\right)$
by applying the bilinear transform. Therefore, a local coordinate
system $\left(u,v\right)$ is introduced \cite{heising1997}. Its'
origin is in the upper left grid point of the quadrilateral in the
current frame, as shown in \fig{\ref{fig:bilinear-transform-patch}}.
The $\left(u,v\right)$ coordinates are computed by subtracting the
upper left corner $\left(x_{c,11},y_{c,11}\right)$ from the coordinates
in the $\left(x,y\right)$ system. The bilinear transform is then
computed according to (\ref{eq:warping_bilinear_u}) and~(\ref{eq:warping_bilinear_v}).

\begin{eqnarray}
u_{r}\left(u_{c},v_{c}\right) & = & a_{11,i}u_{c}v_{c}+a_{12,i}u_{c}+a_{13,i}v_{c}+a_{14,i}\label{eq:warping_bilinear_u}\\
v_{r}\left(u_{c},v_{c}\right) & = & a_{21,i}u_{c}v_{c}+a_{22,i}u_{c}+a_{23,i}v_{c}+a_{24,i}\label{eq:warping_bilinear_v}
\end{eqnarray}

For simplicity, index $i$ is omitted in the following. The calculations
have to be repeated for every quadrilateral. The coefficients $a_{11}$
to $a_{24}$ of the bilinear image transform matrix are computed from
the motion of the grid points of the current quadrilateral.

\begin{eqnarray*}
n_{u}=x_{c,12}-x_{c,11} & \quad & n_{v}=y_{c,21}-y_{c,11}
\end{eqnarray*}
\begin{eqnarray*}
a_{11} & = & u_{r,22}-u_{r,12}+u_{r,11}-u_{r,21}\\
a_{21} & = & v_{r,22}-v_{r,12}+v_{r,11}-v_{r,21}
\end{eqnarray*}
\begin{eqnarray*}
a_{12}=\frac{u_{r,12}-u_{r,11}}{n_{u}} & \quad & a_{22}=\frac{v_{r,12}-v_{r,11}}{n_{v}}\\
a_{13}=\frac{u_{r,21}-u_{r,11}}{n_{u}} & \quad & a_{23}=\frac{v_{r,21}-v_{r,11}}{n_{v}}\\
a_{14}=\frac{u_{r,11}}{n_{u}} & \quad & a_{24}=\frac{v_{r,11}}{n_{v}}
\end{eqnarray*}

The coordinates in the $\left(x,y\right)$ system can then be computed
by moving the origin back, i.e. by adding the upper left corner $\left(x_{c,11},y_{c,11}\right)$
to the $\left(u,v\right)$ coordinates. For the coordinates of the
grid points, the correspondence simplifies to (\ref{eq:mv-GP-x})
and (\ref{eq:mv-GP-y}), where $mv_{x}\left(x_{c},y_{c}\right)$ contains
the $x$~component of the motion vector of the grid points and $mv_{y}\left(x_{c},y_{c}\right)$
the respective $y$~component.
\begin{eqnarray}
x_{r} & = & mv_{x}\left(x_{c},y_{c}\right)+x_{c}\label{eq:mv-GP-x}\\
y_{r} & = & mv_{y}\left(x_{c},y_{c}\right)+y_{c}\label{eq:mv-GP-y}
\end{eqnarray}

\begin{figure}
\begin{center}
\psfragscanon

\psfrag{w}{%
\parbox[t]{3cm}{%
bilinear\\
transform%
}}

\psfrag{i}{%
\parbox[t]{3cm}{%
inverse\\
bilinear\\
transform%
}}

\includegraphics[width=0.9\columnwidth]{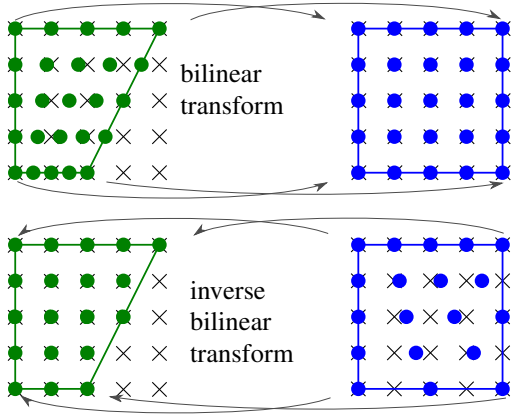}

\psfragscanoff
\end{center}

\protect\caption{\label{fig:mesh-based-compensation-and-inversion}\Meshbased{} compensation
and its inversion in detail showing the source and the destination
of the pixel coordinates. In the upper part, the compensation is shown
according to the bilinear transform of the quadrilateral. In the lower
part, the inverse compensation is shown according to the inverse bilinear
transform of the quadrilateral. Crosses mark positions of the regular
pixel grid where intensity values are available, respectively necessary.
For points beside the regular pixel grid, the corresponding intensities
have to be computed by interpolation.}
\end{figure}

The deformation of every quadrilateral is given by the corresponding
bilinear transform. A detailed example is illustrated in the upper
part of \fig{\ref{eq:L-Haar}}. The bilinear transform of the green
quadrilateral on the left to the blue quadrilateral on the right is
shown. Intensity values at a regular pixel grid are marked by black
crosses. To obtain intensity values for the blue points on the right,
corresponding values are needed from the green positions on the left.
The green positions are computed by the bilinear transform. They are
obtained by interpolation from the available values at the regular
pixel grid.

\subsection{Inversion of the Bilinear Transform}

Within the update step of the \ls{}~(\ref{eq:L-Haar-lifting}),
the compensation has to be inverted \cite{bozinovic2005}. During
the estimation process, movements of the grid points are only allowed
if the corresponding bilinear transform remains invertible. This will
be shown in \sect{(\ref{sub:Restrict-Movement-of-GPs})}. For an invertible
bilinear transform, the inverse can be computed \cite{heising1997}
by solving (\ref{eq:warping_bilinear_u}) and (\ref{eq:warping_bilinear_v})
for $\left(u_{c},v_{c}\right)$. For example, solving (\ref{eq:warping_bilinear_u})
for $u_{c}$ results in 
\begin{eqnarray}
u_{r}-a_{13}v_{c}-a_{14} & = & \left(a_{11}v_{c}+a_{12}\right)u_{c}.\label{eq:warping_bilinear_u_solved_for_x}
\end{eqnarray}
Inserting (\ref{eq:warping_bilinear_u_solved_for_x}) into (\ref{eq:warping_bilinear_v})
and solving for $v_{c}$ results in a quadratic equation (\ref{eq:bilin_warp_inv_y}).

\begin{eqnarray}
\alpha v_{c}^{2}+\beta v_{c}+\gamma & = & 0\label{eq:bilin_warp_inv_y}
\end{eqnarray}
\begin{eqnarray*}
\alpha & = & a_{23}a_{11}-a_{21}a_{13}\\
\beta & = & a_{21}u_{r}-a_{11}v_{r}+a_{24}a_{11}\\
 &  & -a_{21}a_{14}-a_{22}a_{13}+a_{23}a_{12}\\
\gamma & = & a_{22}u_{r}-a_{12}v_{r}+a_{24}a_{12}-a_{22}a_{14}
\end{eqnarray*}
\begin{eqnarray}
v_{c_{1,2}} & = & \frac{-\beta\pm\sqrt{\beta^{2}-4\alpha\gamma}}{2\alpha}\text{, where }\alpha\neq0\label{eq:bilin_warp_inv_soly}
\end{eqnarray}

The desired solution for $v_{c}$ can then be determined by choosing
the one that lies within the quadrilateral in the current frame. For
$\alpha=0$, (\ref{eq:bilin_warp_inv_y}) simplifies to $\beta v_{c}+\gamma=0$.

The inverse of the compensation for the update step is obtained by
computing this mapping for all pixels in all quadrilaterals. Due to
the mapping, pixel values are needed at fractional pixel positions
within the \Hpartband{} $\HP_{t}$. This is illustrated in the lower
part of \fig{\ref{fig:mesh-based-compensation-and-inversion}}. The
blue positions on the right are computed from the green positions
by the inverse bilinear transform. The desired intensity values at
the blue points on the right are computed by interpolation from the
available positions at the regular pixel grid.

In \cite{secker2003}, an approximation of the inversion is done by
negating the motion vectors. The approximation error increases, the
smaller the quadrilaterals and the larger the motion of the grid points
are \cite{schnurrer2012mmsp}. This can lead to an increase of the
energy in the \Hpartband{} and artifacts in the \Lpartband{}~\cite{schnurrer2012mmsp}.

\section{\label{sec:Proposed-Grid-Point-ME}Proposed Grid Point Motion Estimation}

The most challenging part is the estimation of the grid point motion.
Neighboring grid points influence each other. In order to find the
optimum solution, all combinations of motion have to be tested. This
is computationally by far to complex. In \cite{nakaya1994,sullivan1991},
an efficient iterative refinement method is proposed. Independent
grid points of the mesh are organized into sets. In every refinement
iteration, the sets are processed subsequently but all grid points
within one set can be processed in parallel. This is illustrated in
\fig{\ref{fig:mesh-based-compensation-overview}} on the left. Grid
points from one set are marked with the same symbol and can be processed
in parallel. The two grid points within the gray marked quadrilaterals
can be optimized without interfering with each other.

The following subsections explain the different steps of the iterative
refinement of the grid point motion estimation; namely our adaptions
to the used hierarchic estimation \cite{heising2001}, the restriction
of the movement, the proposed regularization, and finally the proposed
metric for finding the optimum update within one refinement step.
Finally a novel metric is introduced to evaluate the smoothness of
the resulting mesh deformation.

\subsection{Hierarchic Estimation}

\begin{figure*}[!t]

\psfragscanon
\psfrag{bs1}{$\meshbs_1$}
\psfrag{bs2}{$\meshbs_2$}
\psfrag{bs3}{$\meshbs_3$}
\psfrag{bs4}{$\meshbs_4$}
\hfill{}\includegraphics[height=3.5cm]{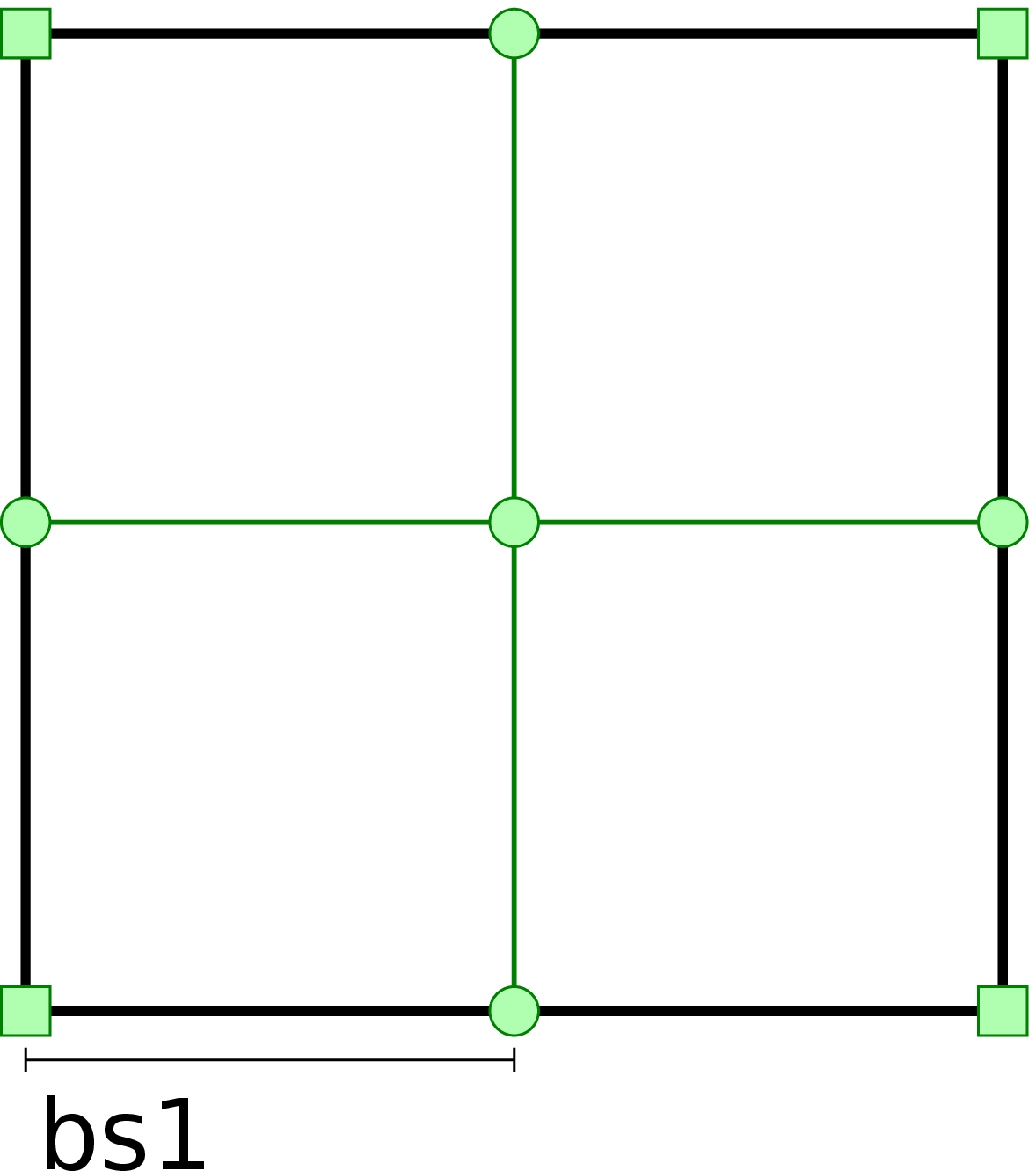}\hfill{}\includegraphics[height=4cm]{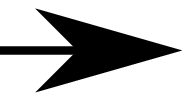}\hfill{}\includegraphics[height=3.5cm]{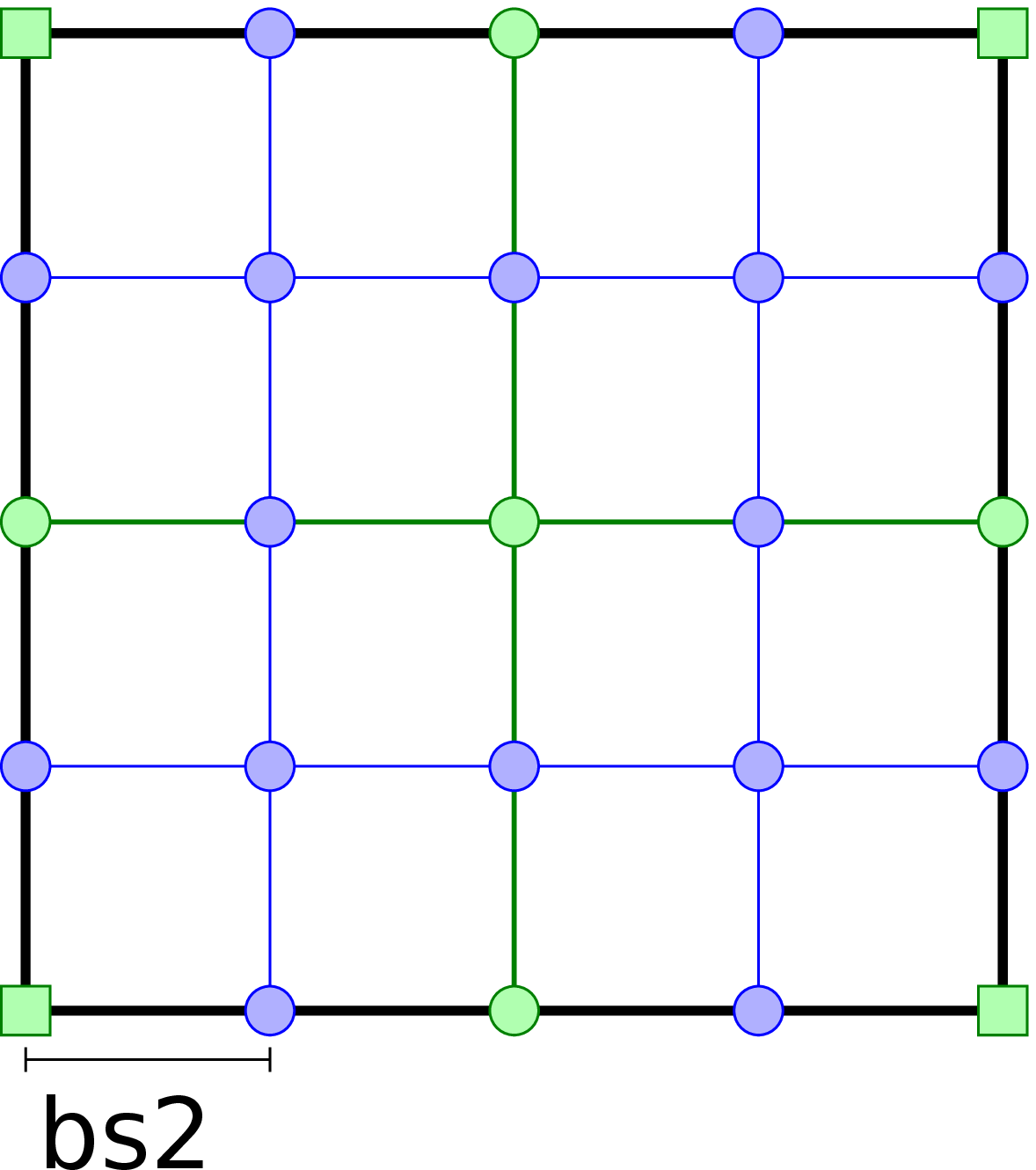}\hfill{}\includegraphics[height=4cm]{img/hierarch_arrow}\hfill{}\includegraphics[height=3.5cm]{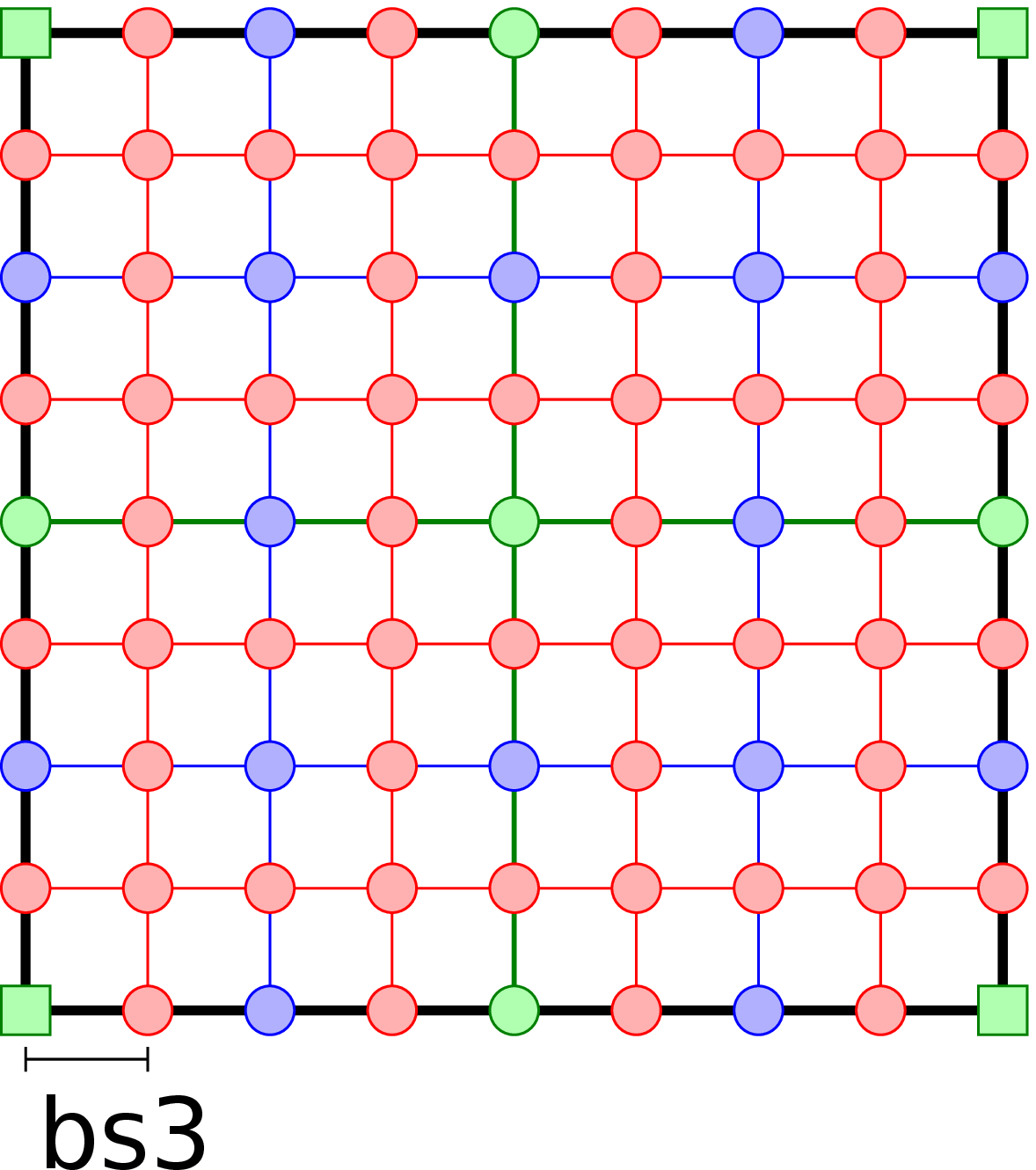}\hfill{}\includegraphics[height=4cm]{img/hierarch_arrow}\hfill{}\includegraphics[height=3.5cm]{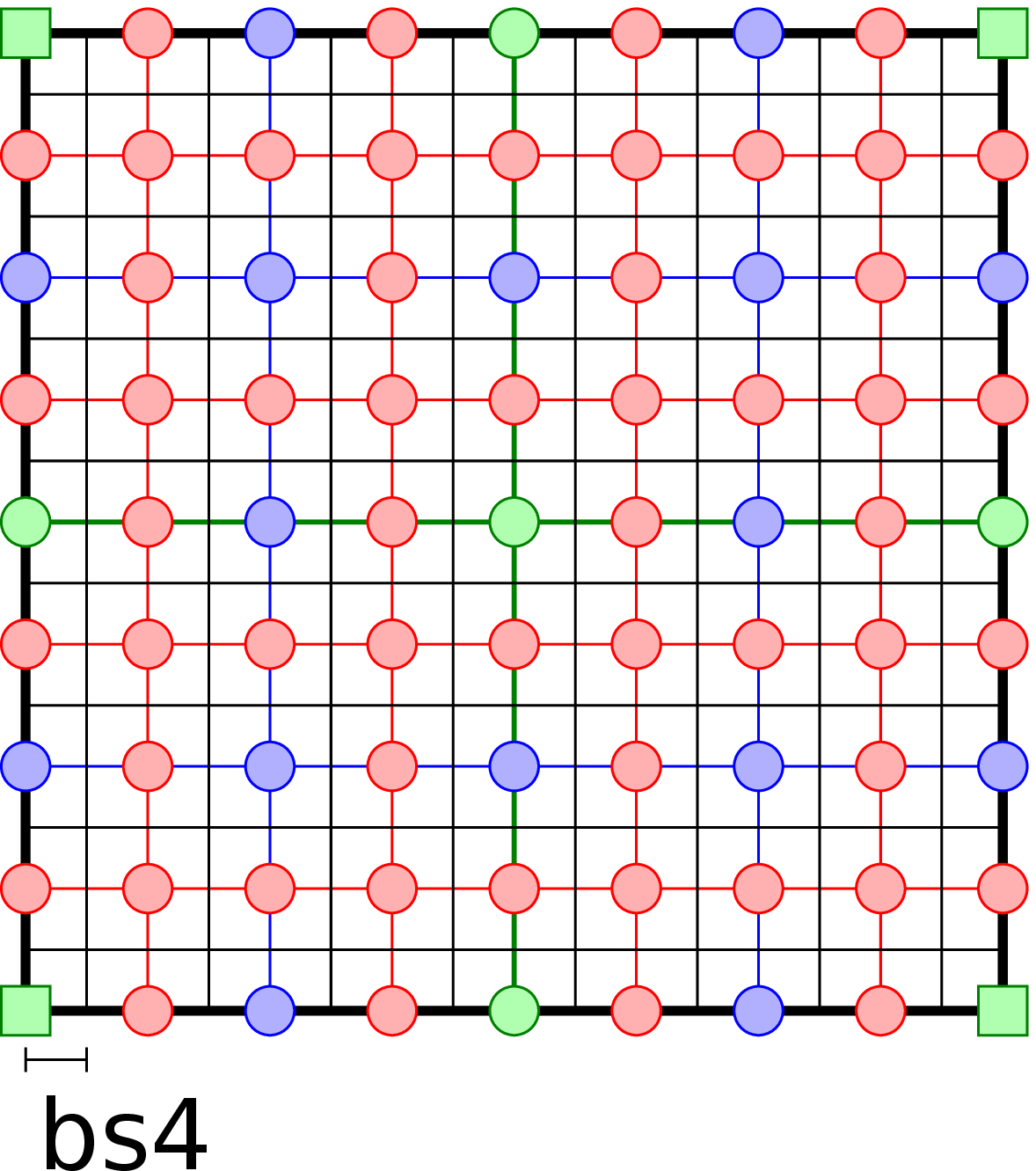}\hfill{}

\psfragscanoff

\protect\caption{\label{fig:hierarchic-estimation}The illustration shows a mesh which
corresponds to the complete frame at different steps of the hierarchic
estimation. The symbols mark the grid points of the mesh. By the very
coarse mesh grid in the beginning on the left, the mesh can better
represent larger moving areas. In every hierarchic step, grid points
are added and the mesh can adapt to smaller local displacements as
well.}
\end{figure*}

With the aim to estimate the movement of larger structures better,
a hierarchic estimation approach can be used \cite{heising2001}.
The deformation of the mesh is estimated with different quadrilateral
sizes, i.e., from coarse to fine mesh resolution. From left to right
in \fig{\ref{fig:hierarchic-estimation}}, the quadrilateral size
is reduced. The first mesh has a quadrilateral size of $\meshbs_{1}$.
After a few iterations, the quadrilateral size is reduced to $\meshbs_{2}$
by adding the blue colored grid points. The motion vectors of the
blue grid points are obtained by bilinear interpolation from the neighboring
green grid points. Subsequently, the motion vectors of all grid points
are refined. Afterwards, the same principle is applied for the red
grid points. With every hierarchy step, more grid points are added
and the quadrilateral size is reduced. Thus, the mesh is able to compensate
finer displacements with every hierarchy step.

In \cite{heising2001}, a \meshbased{} prediction is computed. The
estimation can be accelerated by using a larger refinement step size
$\meshsr$ for the larger mesh size to obtain a faster estimation
\cite{heising2001}. The compensation has to be inverted in the \UStep{}
of the \ls{}. To avoid converging to a minimum further away in periodic
texture structures, we the step size is kept small and constant.

Only the number of grid points is increased during the estimation.
The estimation is performed on the full frame resolution. An additional
reduction of the resolution of the frame during the estimation leads
to worse results, especially when fine detailed structures are omitted
due to downsampling.
 For this reason, the frame resolution was not reduced during the
estimation process.

Another advantage of a hierarchic approach is that movements of structures
that lie within different smaller quadrilaterals can be covered with
a deductive approach by starting with very large quadrilaterals.
Only a few iterations are computed for a small refinement search range
to avoid finding a 'wrong' local minimum for periodic structures.

\subsection{\label{sub:Restrict-Movement-of-GPs}Restrict Movement of the Grid
Points}

There are several reasons for preferring or allowing only certain
updates during the refinement. First of all, the motion compensation
has to be inverted in the update step of the \wt{}. Movements of
the grid points are only allowed when the resulting image transform
remains invertible. 

\begin{figure}
\begin{center}
\psfragscanon
\psfrag{A}{$A$}
\psfrag{B}{$B$}
\psfrag{C}{$C$}
\psfrag{D}{$D$}
\psfrag{E}{$E$}
\psfrag{F}{$F$}
\psfrag{G}{$G$}
\psfrag{H}{$H$}
\psfrag{point}{$P$}
\psfrag{bs}{bs}
\psfrag{sr}{$\meshsr$}

\includegraphics[width=0.95\columnwidth]{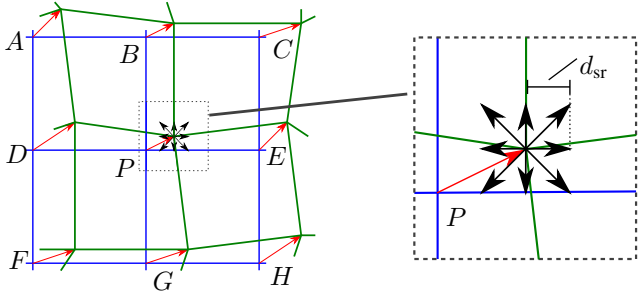}

\psfragscanoff
\end{center}

\protect\caption{\label{fig:estimation_gp_refinement}Refinement for grid point $P$
is considered. The black arrows point to the candidates that are tested
for the refinement of the movement of $P$ with search range $\meshsr$.
On the right, a zoom for the marked region on the left is shown.}

\end{figure}

For explaining the procedure, the refinement of the center grid point
$P$, shown in \fig{\ref{fig:estimation_gp_refinement}}, is considered
in the following.

The determinant of the Jacobian of a bilinear transform (\ref{eq:warping_bilinear_u}),
(\ref{eq:warping_bilinear_v}) can be evaluated as sufficient criterion
for invertibility \cite{glasbey1998review}, \cite{wang1996p1}. The
determinant of the Jacobian (\ref{eq:mesh-determinant}) is dependent
on the image coordinates and forms a plane in 3-D space. The necessary
derivatives can be computed according to (\ref{eq:warping-bilinear-derivatives}).
The bilinear transform is invertible, when the determinant is positive
for every position in a quadrilateral \cite{glasbey1998review}. To
fulfill this, it is sufficient to evaluate the planar equation of
the determinant at the four corner points of a quadrilateral. For
stability reasons, we further add a threshold $\thresDeterJacobian>0$.

\begin{eqnarray}
\det\left(\begin{array}{cc}
\frac{\partial u_{r}}{\partial u_{c}} & \frac{\partial u_{r}}{\partial v_{c}}\\
\frac{\partial v_{r}}{\partial u_{c}} & \frac{\partial v_{r}}{\partial v_{c}}
\end{array}\right) & = & \frac{\partial u_{r}}{\partial u_{c}}\frac{\partial v_{r}}{\partial v_{c}}-\frac{\partial v_{r}}{\partial u_{c}}\frac{\partial u_{r}}{\partial v_{c}}\label{eq:mesh-determinant}\\
 & = & \left(a_{21}a_{11}-a_{21}a_{11}\right)u_{c}v_{c}\nonumber \\
 &  & +\left(a_{12}a_{21}-a_{22}a_{11}\right)u_{c}\nonumber \\
 &  & +\left(a_{23}a_{11}-a_{13}a_{21}\right)v_{c}\nonumber \\
 &  & +\left(a_{12}a_{23}-a_{13}a_{22}\right)\nonumber \\
 & \geq & \thresDeterJacobian\nonumber 
\end{eqnarray}
\begin{align}
\frac{\partial u_{r}\left(u_{c},v_{c}\right)}{\partial u_{c}}=a_{11}v_{c}+a_{12} &  & \frac{\partial u_{r}\left(u_{c},v_{c}\right)}{\partial v_{c}}=a_{11}u_{c}+a_{13}\label{eq:warping-bilinear-derivatives}\\
\frac{\partial v_{r}\left(u_{c},v_{c}\right)}{\partial u_{c}}=a_{21}v_{c}+a_{22} &  & \frac{\partial v_{r}\left(u_{c},v_{c}\right)}{\partial v_{c}}=a_{21}u_{c}+a_{23}\nonumber 
\end{align}

The black arrows in the center of \fig{\ref{fig:estimation_gp_refinement}}
illustrate the refinement positions for the motion vector of grid
point $P$ in the current iteration. For every tested position in
the refinement procedure, for each neighboring quadrilateral of $P$,
namely $ABPD$, $BCEP$, $DPGH$, and $PEHG$, the entries of the
bilinear transform matrix (\ref{eq:warping_bilinear_u}), (\ref{eq:warping_bilinear_v})
have to be computed. Then, for each neighboring quadrilateral, the
determinant of the Jacobian can be evaluated. To maintain the property
for the inversion, a candidate for the refinement is refused if one
of the determinants is smaller than threshold~$\thresDeterJacobian$.

\subsection{Proposed Regularizer for Preferring a Smooth Motion Vector Field}

For preferring a smoother mesh deformation, a regularizer can be
used during the estimation procedure \cite{heising2001}. The more
similar motion vectors of neighboring grid points are, the smoother
the resulting mesh deformation will be. Thereby, the cost function
for finding the optimum candidate in a refinement step is extended
by a regularizer term~$\regterm$ (\ref{eq:regularizer}). This regularizer
term decreases, the more similar the current motion vector candidate
$(mv_{x,\text{cand}},mv_{y,\text{cand}})$ is to the motion vectors
of the neighboring grid points, i.e., the smaller the variance of
the motion vectors \cite{heising2001}. In addition to the literature
\cite{heising2001}, we propose to incorporate the current quadrilateral
size~$\meshbs$ of the hierarchic estimation procedure into the regularizer.
The smaller the quadrilateral size~$\meshbs$, the larger the influence
of the regularizer~$\regterm$ to the cost function. By utilizing
a regularizer, local minima in the estimation are avoided by increasing
the cost for locally very fine-grained deviations of the mesh.

\begin{align}
\regterm(mv_{x,\text{cand}},mv_{y,\text{cand}}) & =\label{eq:regularizer}\\
=\frac{1}{\meshbs}\frac{1}{8}\sum_{\left(x_{\text{GP}},y_{\text{GP}}\right)\in\text{NGP}} & \left(\left(mv_{x}\left(x_{\text{GP}},y_{\text{GP}}\right)-mv_{x,\text{cand}}\right)^{2}+\right.\nonumber \\
+ & \left.\left(mv_{y}\left(x_{\text{GP}},y_{\text{GP}}\right)-mv_{y,\text{cand}}\right)^{2}\right)^{\frac{1}{2}}\nonumber 
\end{align}

An example is given in \fig{\ref{fig:estimation_gp_refinement}}.
The regularizer is computed for every valid candidate, illustrated
by black arrows at the tip of the motion vector of grid point~$P$.
The motion vectors of the eight neighboring grid points $\text{NGP}=\left\{ A,B,C,D,E,F,G,H\right\} $
of the currently considered grid point~$P$ show a smooth movement
in the upper right direction. The cost for a candidate pointing to
the lower right will be higher compared to a candidate pointing in
direction of the neighbors. For a candidate pointing to a different
direction than the neighbors, the similarity metric of the corresponding
deformed quadrilateral has to be small enough to justify this movement.
The next section combines the regularizer with similarity terms and
introduces our proposed cost function for the refinement procedure.

\subsection{Proposed Metric for Estimation}

\begin{figure}
\begin{center}
\psfragscanon
\psfrag{a}{inverse compensation part}
\psfrag{b}{compensation part}

\includegraphics[width=0.9\columnwidth]{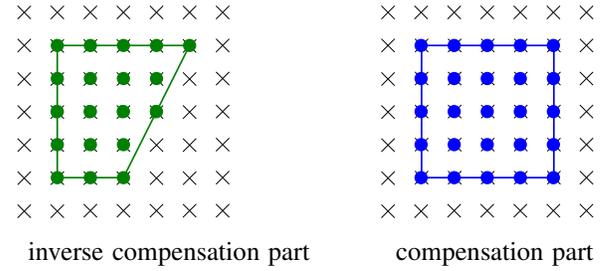}

\psfragscanoff
\end{center}

\protect\caption{\label{fig:components-of-the-metric}For choosing the optimum candidate
during the estimation process, the proposed metric evaluates the similarity
by two parts. For the compensation part, the deformed quadrilateral
of the reference frame is mapped to current frame by applying the
bilinear transform as shown in the upper part of \fig{\ref{fig:mesh-based-compensation-and-inversion}}
and the MSE is computed. The pixels contributing to $\text{MSE}_{\text{comp}}$
are marked on the right. For the inverse compensation part, the quadrilateral
is mapped from the current frame to the reference frame as shown in
the lower part of \fig{\ref{fig:mesh-based-compensation-and-inversion}}
and the MSE is computed again. The pixels contributing to $\text{MSE}_{\text{invcomp}}$
are marked on the right.}
\end{figure}

To determine an optimum candidate in the refinement procedure, a cost
function is utilized. So far~\cite{heising2001}, the cost function
only consists of a similarity term evaluating the prediction error
and a regularizer. The origin of the similarity term is illustrated
in \fig{\ref{fig:components-of-the-metric}} on the right. The bilinear
transform results in a prediction for the blue colored quadrilateral
on the right. For the blue points, the mean squared error $\text{MSE}_{\text{comp}}$
is computed and used for the similarity term.

So far, the state of the art metric $\meshmetricold$ is used where
$\reglambda$ controls the influence of the regularizer~$\regterm$.

\begin{equation}
\meshmetricold=\text{MSE}_{\text{comp}}+\reglambda\regterm\label{eq:metric11}
\end{equation}

The inverse compensation is not considered so far, i.e., how well
the inverse transform of the blue quadrilateral matches the green
quadrilateral. This is illustrated by the green points on the left
of \fig{\ref{fig:components-of-the-metric}}. To incorporate the quality
of the inversion into the metric, the mean squared error $\text{MSE}_{\text{invcomp}}$
for the green points on the left is computed.

The proposed metric (\ref{eq:metric13}) is computed by 

\begin{equation}
\meshmetricprop=\text{MSE}_{\text{comp}}+\text{MSE}_{\text{invcomp}}+\reglambda\regterm.\label{eq:metric13}
\end{equation}

The normalization of the sum of squared differences to the number
of elements is necessary to address the different number of elements
for the compensation and inverse compensation part, shown by the different
number of blue and green points in \fig{\ref{fig:components-of-the-metric}}.
To address this, the mean squared error is used. With this metric,
the optimum candidate is determined during the iterative refinement
procedure.

\subsection{Proposed Smoothness Evaluation of the Mesh Deformation}

In order to determine the smoothness of the mesh with one value,
a metric was developed. The smoothness of the deformation is represented
by $\meshsmoothness\in\left(0;1\right]$, where values closer to 1~correspond
to a shape more similar to a square.

For the green deformed quadrilateral $ABPD$ shown in \fig{\ref{fig:estimation_gp_refinement}},
$\meshsmoothness_{ABPD}$ is computed according to
\begin{equation}
\meshsmoothness_{ABPD}=\frac{\min\left(\overline{AB},\overline{BP},\overline{PD},\overline{DA}\right)}{\max\left(\overline{AB},\overline{BP},\overline{PD},\overline{DA}\right)}\cdot\frac{\min\left(\overline{AP},\overline{BD}\right)}{\max\left(\overline{AP},\overline{BD}\right)}.\label{eq:meshsmoothness}
\end{equation}
The first fraction evaluates the ratio between the minimum and the
maximum length of the edges. The second fraction evaluates the respective
ratio for the diagonals of the quadrilateral. 
\begin{figure}
\begin{center}
\psfragscanon
\psfrag{A}{$A$}
\psfrag{B}{$B$}
\psfrag{C}{$C$}
\psfrag{D}{$D$}
\psfrag{E}{$E$}
\psfrag{F}{$F$}
\psfrag{G}{$G$}
\psfrag{H}{$H$}
\psfrag{point}{$P$}
\psfrag{bs}{bs}
\psfrag{sr}{$\meshsr$}

\psfrag{v1}{$1\cdot1=1$}
\psfrag{v2}{$\frac{1}{3}\frac{\sqrt{5}}{\sqrt{13}}\approx0.2$}
\psfrag{v3}{$1\cdot1=1$}
\psfrag{v4}{$\frac{1}{2}\cdot1=\frac{1}{2}$}
\psfrag{v5}{$1\cdot1=1$}

\includegraphics[width=0.95\columnwidth]{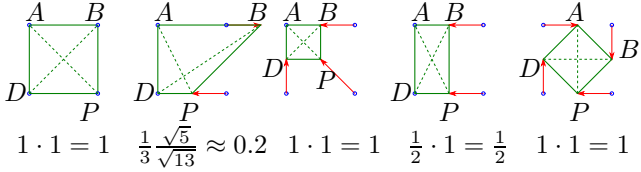}

\psfragscanoff
\end{center}

\protect\caption{\label{fig:mesh-smoothness}Examples for the proposed evaluation of
the mesh smoothness and the corresponding values of $\meshsmoothness_{ABPD}$. }
\end{figure}
\fig{\ref{fig:mesh-smoothness}} shows examples for the smoothness
value of different deformations. The smoothness value becomes smaller
with increasing fine-granular local mesh deformation. Therefore, a
value closer to 1 is desired. This value is computed for all quadrilaterals
of the mesh. To obtain one meaningful value for the smoothness of
the complete mesh, the smoothness value $\meshsmoothness$ of the
complete mesh is computed by averaging the results from all quadrilaterals.
On average, a value closer to 1 corresponds to a smoother deformation
of the complete mesh. This metric is used to evaluate the smoothness
of the mesh deformation after the iterative refinement procedure.

\section{\label{sec:Simulation-Results}Simulation Results}

\begin{table}
\protect\caption{\label{tab:Dynamic-medical-volumes}Dynamic medical volumes used for
simulation}

 \setlength\tabcolsep{4pt} 

\begin{center}
\begin{tabular}{c|c|c|c|c|c|c|c|c|c|c}
\toprule
  & CT & \multicolumn{9}{c}{MR}\tabularnewline
\midrule  
x & 512 & 192 & 192 & 192 & 156 & 192 & 192 & 156 & 156 & 156\tabularnewline 

y & 512 & 256 & 256 & 256 & 192 & 156 & 156 & 192 & 192 & 192\tabularnewline 

z & 128 & 1 & 1 & 1 & 1 & 1 & 1 & 1 & 1 & 1\tabularnewline

t & 10 & 60 & 60 & 60 & 24 & 24 & 24 & 10 & 10 & 10\tabularnewline
\bottomrule
 \end{tabular}
\end{center}
\end{table}

For the simulation, we used dynamic volumes from computed tomography
(CT\footnote{The CT volume data set was kindly provided by Siemens Healthcare.})
and magnetic resonance tomography (MR\footnote{The MR volume data sets were kindly provided by doctor Jacues Beckmann,
radiologist at Katharinen Hospital, Unna.}) showing sequences of the beating heart. \Tab{\ref{tab:Dynamic-medical-volumes}}
lists details for the spatial and temporal resolution. The intensity
values have a co-domain of 12~bit. The spatial resolution of the
CT volumes is typically higher. To be independent of the actual co-domain,
a scaling to $[0;1]$ is done for the estimation process of the mesh
deformation. To evaluate our proposed metric, we perform one compensated
wavelet decomposition step in temporal direction. For the 3-D+t CT
volume, we took the 10~time steps at every position in $z$-direction
as sequence. This results in 128 CT sequences.

\begin{figure*}
 \setlength\tabcolsep{4pt} 

\begin{tabular}{cccc}
\begin{turn}{90}
~
\end{turn} & \includegraphics[width=0.31\textwidth]{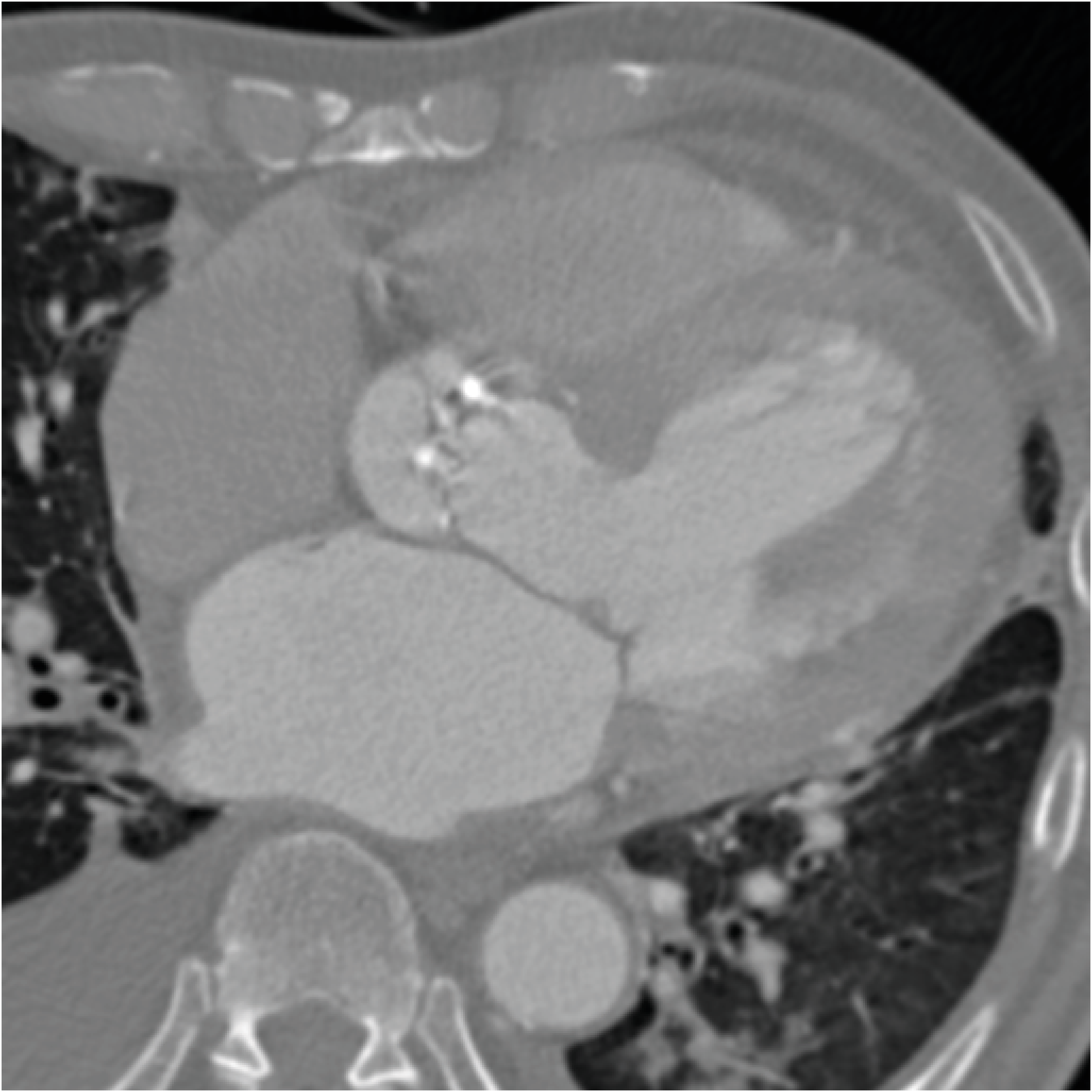} & \includegraphics[width=0.31\textwidth]{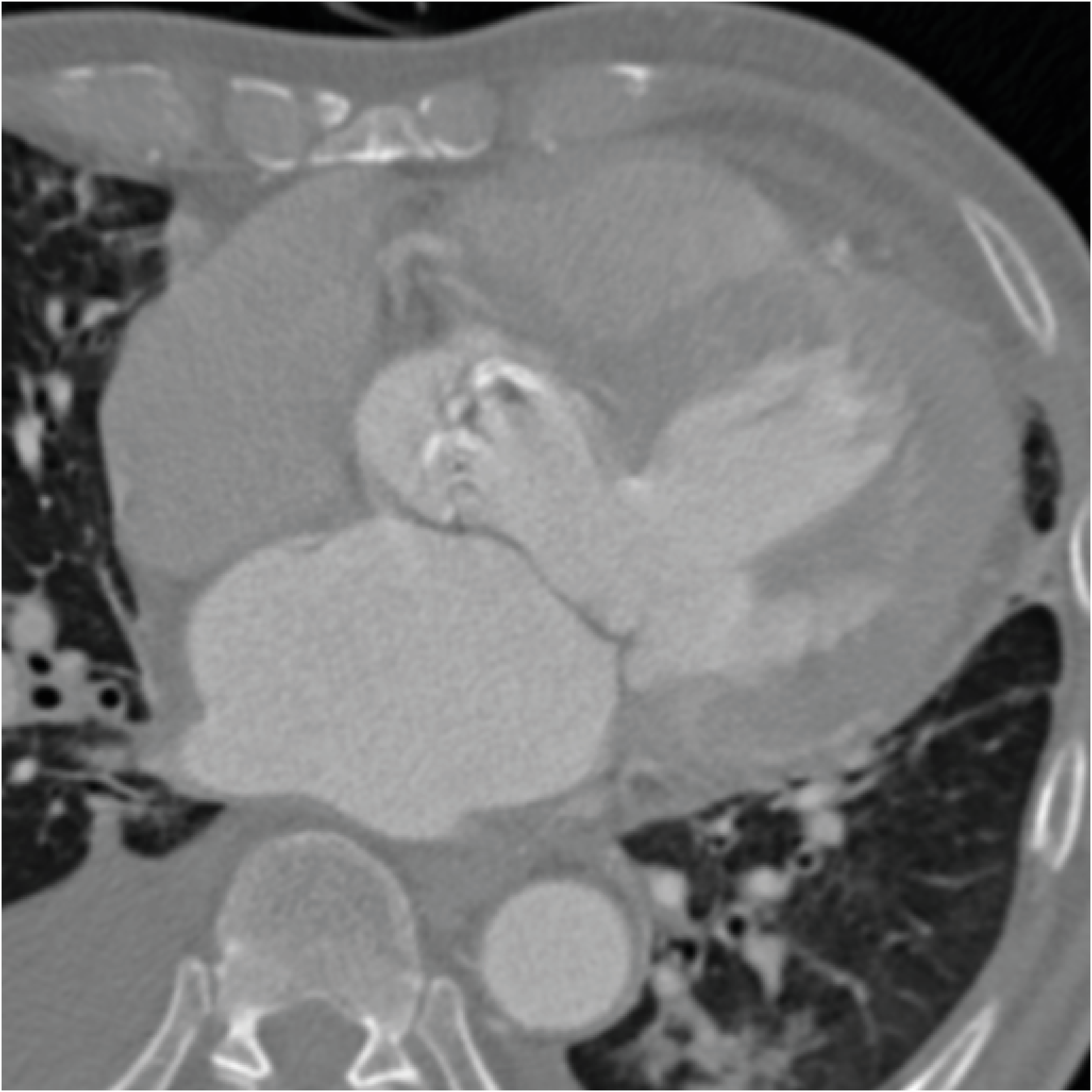} & \includegraphics[width=0.31\textwidth]{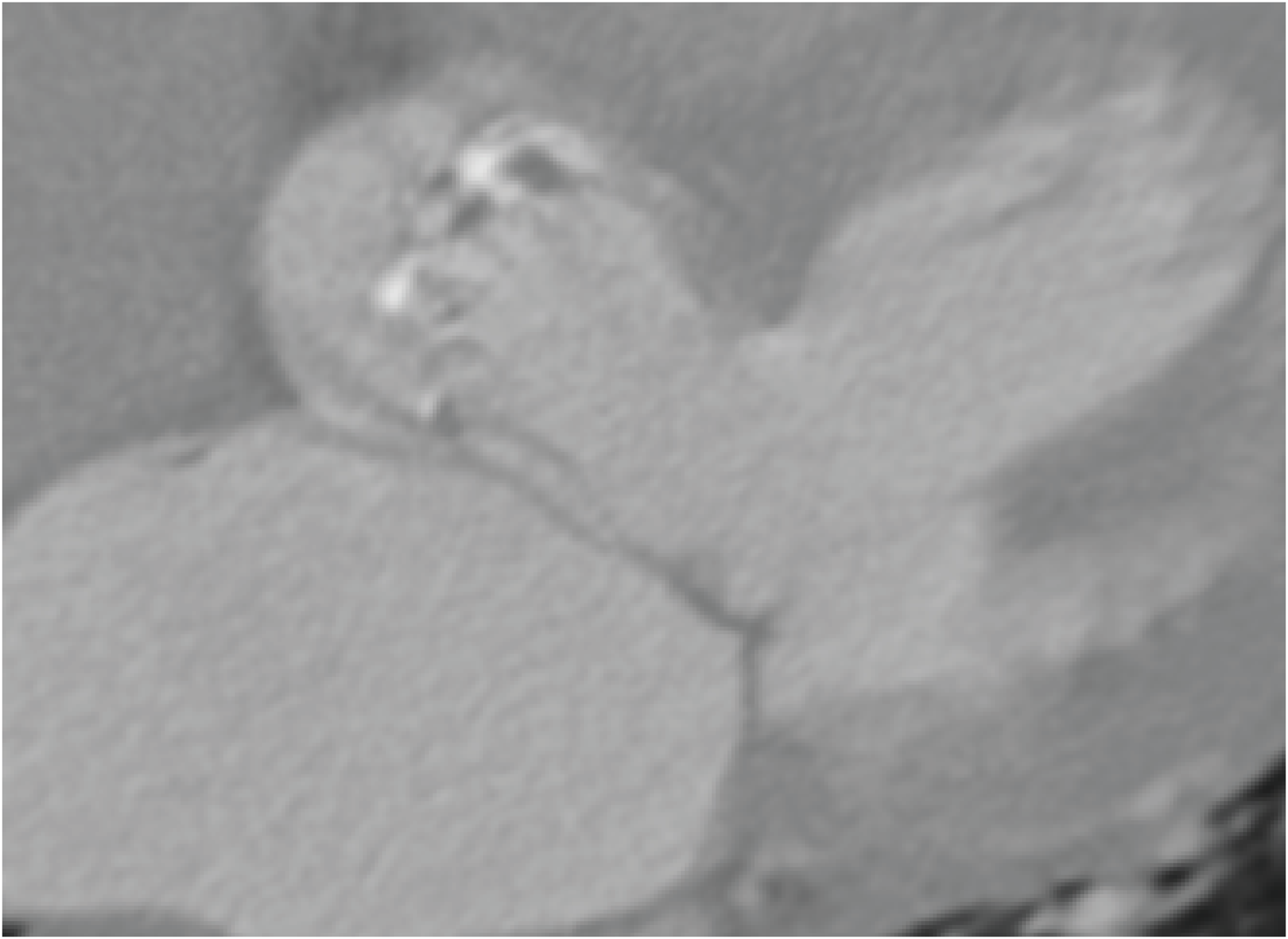}\tabularnewline
 & reference frame $\sigf_{1}$ & current frame $\sigf_{2}$ & $\LP_{1}$(detail), without compensation\tabularnewline
 &  &  & PSNR$\left(\sigf_{1},\LP_{1}\right)=$~\input{img/comparisson/val/vol2_t1-2_z51_m11_LP_hl0_nh_psnr_ref_LP_haar.txt}\tabularnewline
 &  &  & PSNR$\left(\sigf_{2},\LP_{1}\right)=$~\input{img/comparisson/val/vol2_t1-2_z51_m11_LP_hl0_nh_psnr_cur_LP_haar.txt}\tabularnewline
\end{tabular}

\protect\caption{\label{fig:Visual-orig-haar}Visual example for two successive original
frames at slice $z=51$ of the CT volume. From left to right, the
reference frame $f_{1}$, the current frame $f_{2}$, and a detail
of the corresponding frame $\LP_{1}$ of the \Lpartband{} using a
Haar transform without compensation is shown. }
\end{figure*}

\Fig{\ref{fig:Visual-orig-haar}} shows two successive frames of the
CT volume at slice position $z=51$. A detail of the \Lpartband{}
of the Haar transform without compensation is shown on the right.
Due to the large displacement from $\sigf_{1}$ to $\sigf_{2}$, the
\Lpartband{} gets blurred and most details are lost. The separated
structures in the original frames are mixed up and cannot be distinguished
anymore. This example visualizes the need for feasible compensation
methods.

For the \meshbased{} compensation, the threshold for the determinant
of the Jacobian is set to $\thresDeterJacobian=0.2$. Thereby, the
invertibility of the compensation is maintained. Geometrically, this
means, the quadrilaterals remain convex and do not diverge, e.g.,
to triangles \cite{wang1996p1}. \Tab{\ref{tab:iterations-hirarchic-estimation}}
lists the parameters used for the iterative hierarchic estimation
procedure for the CT and the MR volumes. Intensity values at non-integer
positions are calculated by bilinear interpolation. For the CT volumes,
the hierarchic estimation starts with 3~iterations using a search
range $\meshsr=1$ at a quadrilateral size of $\meshbs=256$. In the
next hierarchic step 3~iterations are computed at $\meshbs=128$,
etc. After the second hierarchic step, 6~iterations are computed.
The row 'sum' counts the total number of iterations. For the resolutions
of the MR volumes, a maximum quadrilateral size of $\meshbs=64$ is
reasonable to start the hierarchic estimation with a large grid size
where the quadrilaterals have about the same size and $\meshbs$ is
a power of~2. The last two columns of \Tab{\ref{tab:iterations-hirarchic-estimation}}
correspond to the extension where the influence of a subpixel search
range $\meshsr$  was tested.

\begin{table}
\protect\caption{\label{tab:iterations-hirarchic-estimation}Details for the hierarchic
steps used in the simulation.}
 \setlength\tabcolsep{4pt} 

\begin{center}
\begin{tabular}{c|c|c|c|c|c|c|c||c|c}
\toprule
\multicolumn{2}{c|}{ } & \multicolumn{6}{c||}{hierarchic estimation}& \multicolumn{2}{c}{subpixel} \tabularnewline 
\multicolumn{2}{c|}{ } & \multicolumn{6}{c||}{ }& \multicolumn{2}{c}{extension} \tabularnewline 

\midrule
\multirow{4}{*}{\begin{turn}{90} CT \end{turn}} & $\meshbs$ & 256 & 128 & 64 & 32 & 16 & 8 & 8 & 8\tabularnewline 
 & $\meshsr$ & 1 & 1 & 1 & 1 & 1 & 1 & 0.5 & 0.25 \tabularnewline 
 & \#iter & 3 & 3 & 4 & 5 & 6 & 9 & 1 & 1\tabularnewline
 & sum &  & 6 & 10 & 15 & 21 & 30 & 31 & 32\tabularnewline 
\midrule
\multirow{4}{*}{\begin{turn}{90} MR \end{turn}} & $\meshbs$ & & & 64 & 32 & 16 & 8 & 8 & 8\tabularnewline 
 & $\meshsr$ & & &  1 & 1 & 1 & 1 & 0.5 & 0.25 \tabularnewline 
 & \#iter & & & 10 & 5 & 6 & 9 & 1 & 1\tabularnewline
 & sum & & & & 15 & 21 & 30 & 31 & 32\tabularnewline 
\bottomrule
 \end{tabular}
\end{center}
\end{table}

\begin{figure*}
\begin{center}

 \setlength\tabcolsep{4pt} 

\begin{tabular}{l|l|l}
\multicolumn{2}{c|}{state of the art metric metric $\meshmetricold$} & \multicolumn{1}{c}{proposed metric $\meshmetricprop$}\tabularnewline
\multicolumn{1}{c|}{non hierarchic estimation} & \multicolumn{2}{c}{hierarchic estimation}\tabularnewline
PSNR$\left(\sigf_{1},\LP_{1}\right)=$~\input{img/comparisson/val/vol2_t1-2_z51_m11_LP_hl4_nh_psnr_ref_LP.txt} & PSNR$\left(\sigf_{1},\LP_{1}\right)=$~\input{img/comparisson/val/vol2_t1-2_z51_m11_LP_hl4_hier_psnr_ref_LP.txt} & PSNR$\left(\sigf_{1},\LP_{1}\right)=$~\input{img/comparisson/val/vol2_t1-2_z51_m13_LP_hl4_hier_psnr_ref_LP.txt}\tabularnewline
PSNR$\left(\sigf_{2},\mathcal{W}_{1\rightarrow2}\left(\LP_{1}\right)\right)=$~\input{img/comparisson/val/vol2_t1-2_z51_m11_LP_hl4_nh_psnr_cur_LPMC.txt} & PSNR$\left(\sigf_{2},\mathcal{W}_{1\rightarrow2}\left(\LP_{1}\right)\right)=$~\input{img/comparisson/val/vol2_t1-2_z51_m11_LP_hl4_hier_psnr_cur_LPMC.txt} & PSNR$\left(\sigf_{2},\mathcal{W}_{1\rightarrow2}\left(\LP_{1}\right)\right)=$~\input{img/comparisson/val/vol2_t1-2_z51_m13_LP_hl4_hier_psnr_cur_LPMC.txt}\tabularnewline
\multicolumn{1}{c}{\includegraphics[width=0.31\textwidth]{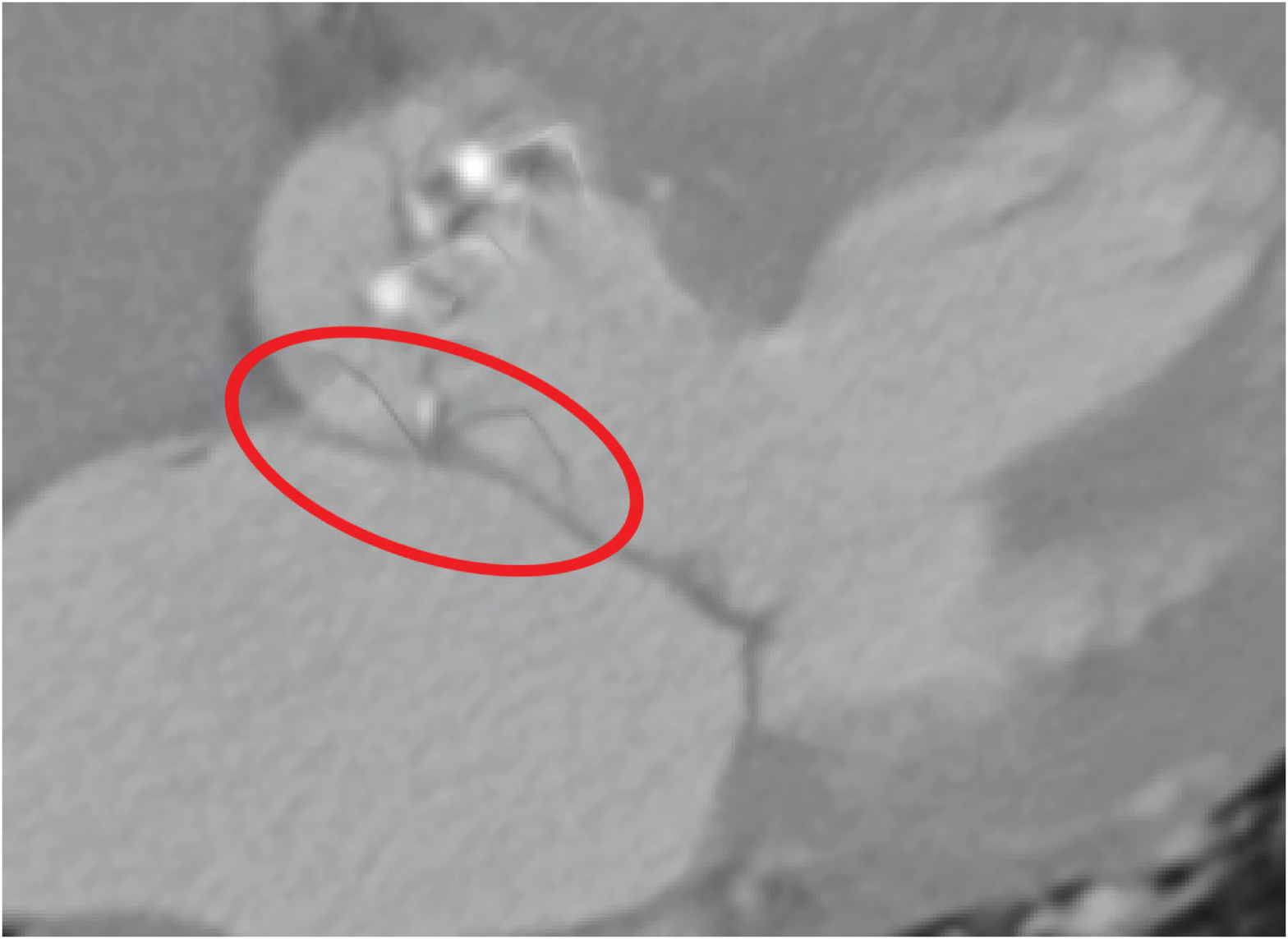}} & \multicolumn{1}{c}{\includegraphics[width=0.31\textwidth]{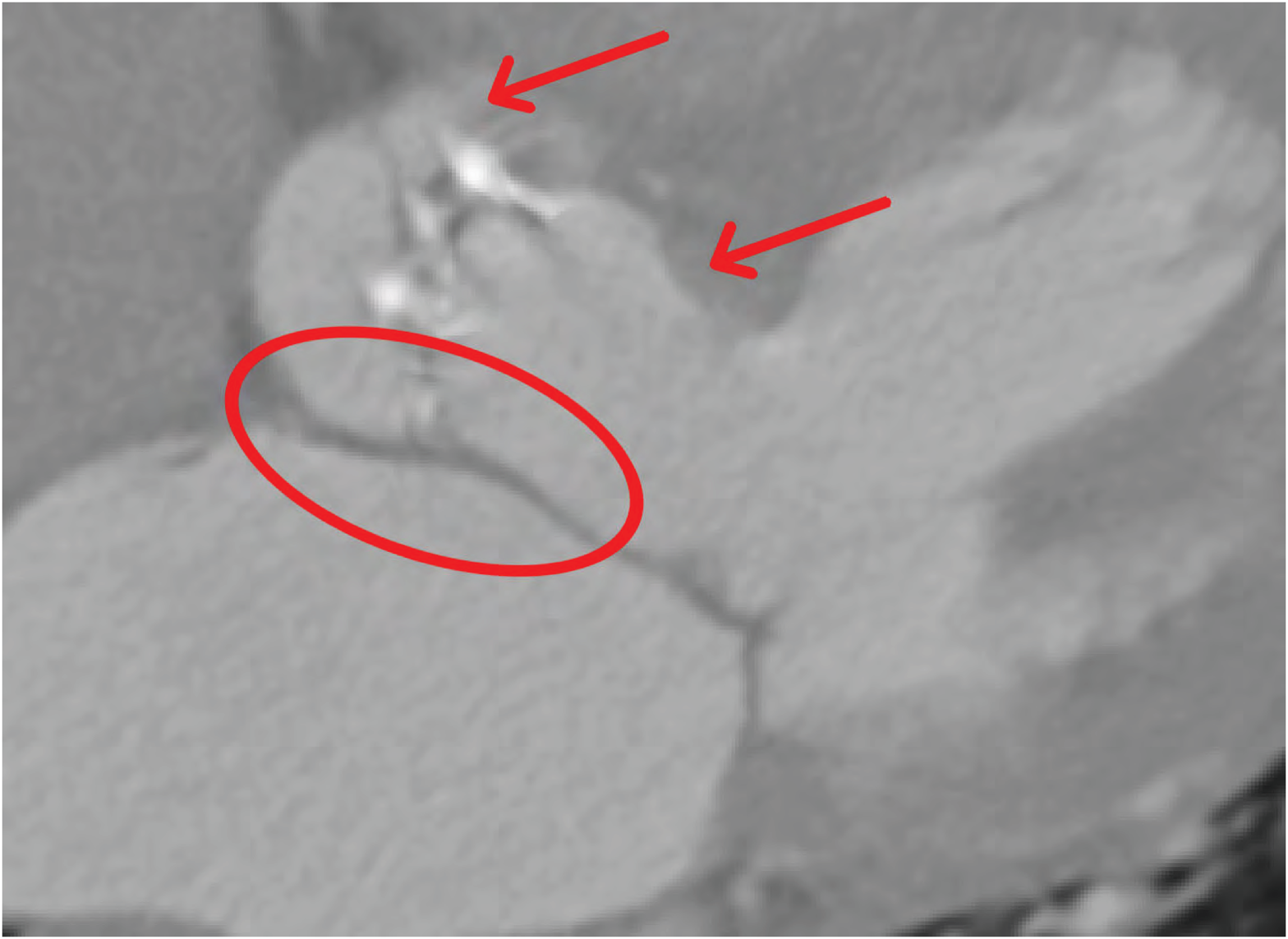}} & \multicolumn{1}{c}{\includegraphics[width=0.31\textwidth]{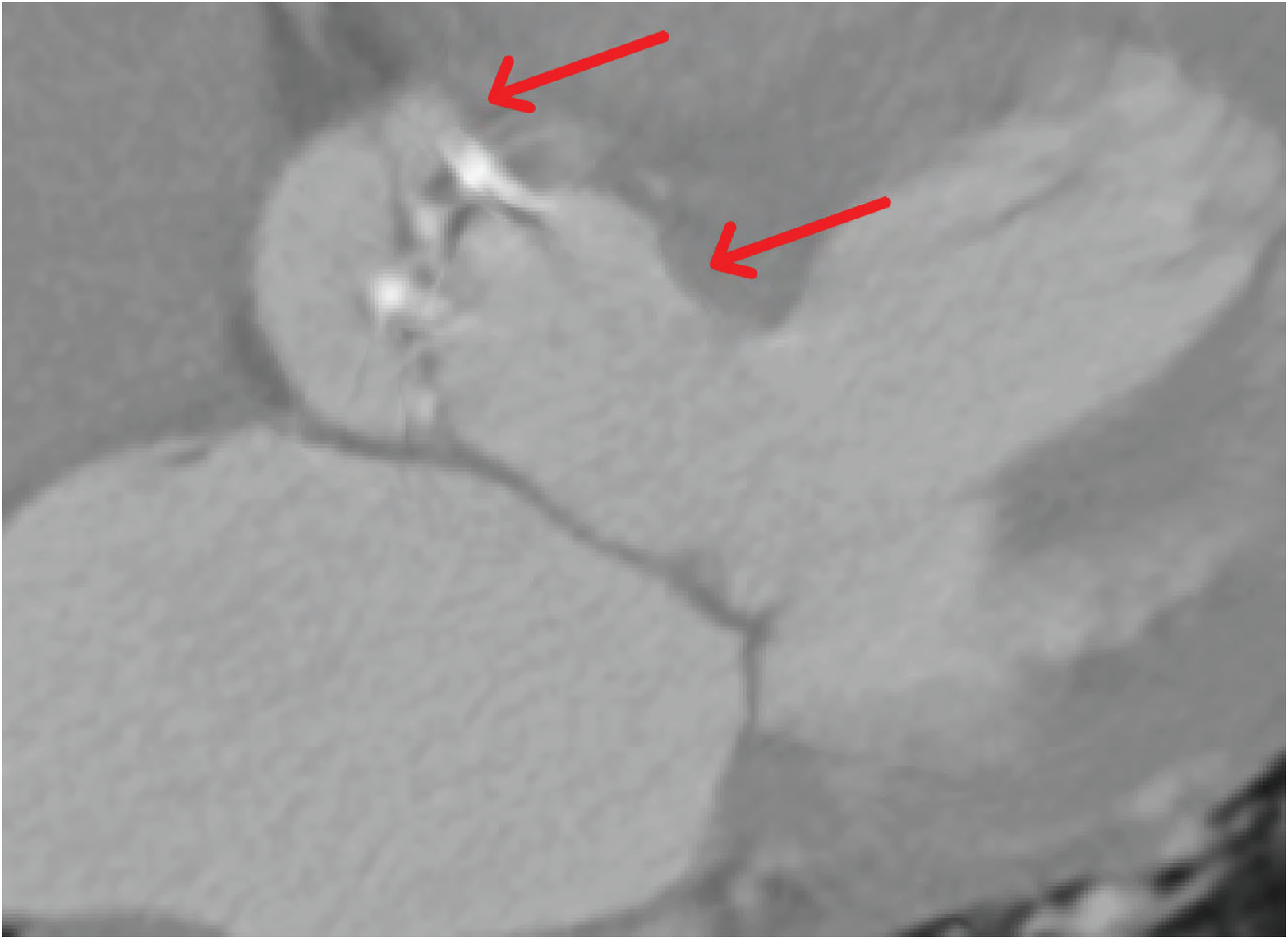}}\tabularnewline
\end{tabular}

\end{center}

\protect\caption{\label{fig:visual-comparison-lowpass}Visual comparison of the resulting
\Lpartbands{} from a mesh compensated \wt{}. The regularizer parameter
was set to $\reglambda=0.0004$. On the left, the state of the art
metric $\meshmetricold$ without hierarchic estimation was used. In
the center, the result is shown using the state of the art metric
and hierarchic estimation. On the right, the result of our proposed
metric $\meshmetricprop$ with hierarchic estimation is shown. For
the proposed metric, the corresponding mesh is shown on the bottom
of in \fig{\ref{fig:visual-comparisson-mesh}}. From left to right,
the artifacts marked in red disappear.}
\end{figure*}

\begin{figure}
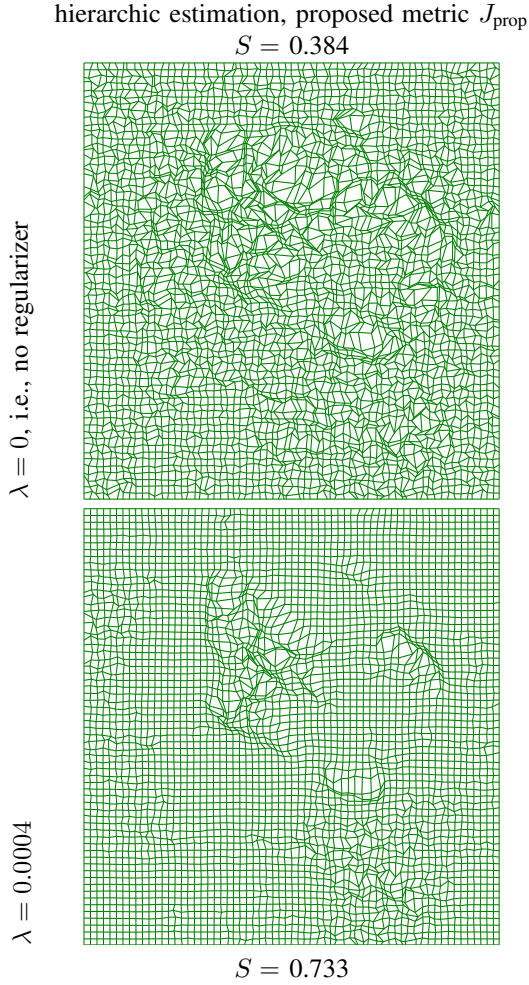

    \setlength\figureheight{0.32\textwidth}     \setlength\figurewidth{0.32\textwidth}   

 \setlength\tabcolsep{4pt} 

\begin{center}

\begin{tabular}{cc}
 & hierarchic estimation, proposed metric $\meshmetricprop$\tabularnewline
 & $\meshsmoothness=$~\input{img/comparisson/val/vol2_t1-2_z51_m13_LP_hl0_hier_mesh_smooth_comb.txt}\tabularnewline
\begin{turn}{90}
$\reglambda=0$, i.e., no regularizer
\end{turn} & \input{img/comparisson/vol2_t1-2_z51_m13_mesh_hl0_hier.tex}\tabularnewline
\begin{turn}{90}
$\reglambda=0.0004$
\end{turn} & \input{img/comparisson/vol2_t1-2_z51_m13_mesh_hl4_hier.tex}\tabularnewline
 & $\meshsmoothness=$~\input{img/comparisson/val/vol2_t1-2_z51_m13_LP_hl4_hier_mesh_smooth_comb.txt}\tabularnewline
\end{tabular}

\end{center}

\protect\caption{\label{fig:visual-comparisson-mesh}Comparison of the influence of
the regularizer on the resulting mesh. Our proposed metric $\meshmetricold$
and hierarchic estimation was used. The upper mesh results from the
estimation without regularization. For the lower mesh, the regularizer
parameter was set so $\reglambda=0.0004$ and a detail of the corresponding
frame of the \Lpartband{} is shown in \fig{\ref{fig:visual-comparison-lowpass}}
on the right. }
\end{figure}

\begin{table*}[!p]
\protect\caption{\label{tab:PSNR-in-dB}PSNR in dB, Mesh smoothness $\meshsmoothness$,
and rate for the mesh parameters for different values of the parameter
$\reglambda$ for the regularizer}

 \setlength\tabcolsep{4pt} 

\begin{center}
\begin{tabular}{c|c|rrrrr|rrrrr}
\toprule
 & & \multicolumn{5}{c|}{state of the art metrik $\meshmetricold$} &\multicolumn{5}{c}{proposed metrik $\meshmetricprop$} \tabularnewline

 & & $\reglambda=0$ & $0.0003$ & $0.0004$& $0.0005$& $0.0007$ & $\reglambda=0$& $0.0003$ & $0.0004$& $0.0005$& $0.0007$ \tabularnewline

\midrule
\multirow{4}{*}{\begin{turn}{90} CT \end{turn}}

 & $\overline{\text{PSNR}\left(\sigf_{2t-1},\LP_{t}\right)}$ &  49.28 & 48.04 & 47.85 & 47.70 & 47.44 & 49.70 & 48.66 & 48.48 & 48.33 & 48.09 \tabularnewline

 & $\overline{\text{PSNR}\left(\sigf_{2t},\mathcal{W}_{2t-1\rightarrow2t}\left(\LP_{t}\right)\right)}$ &  48.56 & 47.45 & 47.28 & 47.13 & 46.90 & 48.38 & 47.73 & 47.60 & 47.49 & 47.31  \tabularnewline

 & smoothness $\meshsmoothness$ &  0.393 & 0.856 & 0.878 & 0.893 & 0.913 & 0.397 & 0.791 & 0.820 & 0.839 & 0.866 \tabularnewline

 & filesize in kbyte &  5.35 & 2.12 & 1.98 & 1.88 & 1.76 & 5.33 & 2.55 & 2.36 & 2.23 & 2.05 \tabularnewline


\multirow{4}{*}{\begin{turn}{90} MR \end{turn}}
 & $\overline{\text{PSNR}\left(\sigf_{2t-1},\LP_{t}\right)}$ &   49.98 & 48.86 & 48.70 & 48.56 & 48.36 & 50.02 & 49.26 & 49.13 & 49.00 & 48.82 \tabularnewline

 & $\overline{\text{PSNR}\left(\sigf_{2t},\mathcal{W}_{2t-1\rightarrow2t}\left(\LP_{t}\right)\right)}$ &   48.31 & 48.28 & 48.22 & 48.17 & 48.12 & 48.23 & 48.36 & 48.32 & 48.28 & 48.23 \tabularnewline

 & smoothness $\meshsmoothness$ &   0.693 & 0.913 & 0.921 & 0.927 & 0.934 & 0.697 & 0.885 & 0.897 & 0.906 & 0.916 \tabularnewline

 & filesize in kbyte &   0.48 & 0.27 & 0.26 & 0.25 & 0.24 & 0.48 & 0.30 & 0.29 & 0.28 & 0.26 

\end{tabular}
\end{center}
\end{table*}

For the example given in \fig{\ref{fig:Visual-orig-haar}}, \fig{\ref{fig:visual-comparison-lowpass}}
shows details of the \Lpartband{} from a compensated Haar transform
using different parameters for the mesh.   The regularizer parameter
was set to $\reglambda=0.0004$. On the left, no hierarchic estimation
was used, i.e. 30~iterations with $\meshsr=1$ were computed at grid
size $\meshbs=8$. In the last two columns, hierarchic estimation
according to \Tab{\ref{tab:iterations-hirarchic-estimation}} was
applied. In the first two columns, the state of the art metric $\meshmetricold$
and in the last column our proposed metric $\meshmetricprop$ was
used during the estimation process. The given PSNR values show the
similarity of the \Lpartband{} frames to the original frames. Usually,
only the similarity between the \Lpartband{} frame and the corresponding
original frame is considered by evaluating PSNR$\left(\sigf_{1},\LP_{1}\right)$.
However, the \Lpartband{} is a representation for the complete original
volume. To evaluate additionally the similarity to the current frame,
PSNR$\left(\sigf_{2},\mathcal{W}_{1\rightarrow2}\left(\LP_{1}\right)\right)$
is evaluated. Therefore, the \Lpartband{} frame $\LP_{1}$ is warped
to the time step of the current frame $\mathcal{W}_{1\rightarrow2}$.

Without the hierarchic estimation, visual artifacts are clearly visible
in left image of \fig{\ref{fig:visual-comparison-lowpass}}. The
sequence contains displacements which are too large to be considered
by the small quadrilaterals of the non hierarchic approach. The convergence
of the mesh to local minima results in disturbing artifacts in the
\Lpartband{} showing structures which have not been in the original
volume. By starting the estimation with very large quadrilaterals,
larger and global displacements can be modeled better. By using the
hierarchic estimation, the artifacts marked by the red ellipse mostly
disappear in the center image of \fig{\ref{fig:visual-comparison-lowpass}}.
However, some artifacts and blurriness remain in the center image
visible in the area pointed by the red arrows. Especially the light
gray structure is sharper in the right image. Although the bilinear
transform is inverted, artifacts can occur when only one direction
of the image transform is considered. The detail shown on the right
results from our proposed metric and contains sharper structures.
Compared to the state of the art metric used in the center, the visual
quality is improved significantly.

\fig{\ref{fig:visual-comparisson-mesh}} shows the impact of the
regularizer. The upper mesh results from the estimation without regularization.
Fine granular local deviations of the quadrilaterals are visible.
Due to noise in the original volume, the mesh is deformed even in
mostly flat areas without displacement. By incorporating a feasible
regularizer, the resulting deformation of the mesh is smoother and
corresponds more to the assumptions for the displacement within dynamic
volumes. Solely the combination of regularizer and hierarchic estimation
yield to a reasonable estimation. The lower mesh in \fig{\ref{fig:visual-comparisson-mesh}}
corresponds to the detail shown on the right of \fig{\ref{fig:visual-comparison-lowpass}}.

The results for the CT and the MR volumes are very similar, hence
the average for all volumes is given in \tab{\ref{tab:PSNR-in-dB}}.
The table shows that by incorporating a regularizer $\left(\reglambda>0\right)$,
the PSNR results become a little smaller. Basically, the regularizer
modifies the cost function in the estimation process. Deformations
which might be optimum with respect to the squared error are made
more expensive in case of causing a non smooth mesh deformation. The
strength of the regularizer can be controlled using the parameter
$\reglambda$. For $\reglambda=0$, the regularizer has no influence
on the determination of the grid point movement. For $\reglambda>0$,
the restriction on the grid point movement increases with an increasing
value of $\reglambda$. Deviations due to noise in smooth regions
can be avoided by using a reasonable value of $\reglambda$. If the
value is chosen to high, the mesh cannot compensate the displacement
anymore.

From left to right, \tab{\ref{tab:PSNR-in-dB}} lists the results
obtained with \meshbased{} compensation using the state of the art
metric $\meshmetricold$ and our proposed metric $\meshmetricprop$
with increasing $\reglambda$. As can be seen from the results in
\tab{\ref{tab:PSNR-in-dB}}, turning off the regularizer $\left(\reglambda=0\right)$
leads to a significant decrease in the mesh smoothness and a significant
increase in the rate needed for the mesh parameters. The rate is obtained
using \cite{fowler2000qccpack}. $\meshmetricold$ only optimizes
the compensation direction. The decrease of the similarity of the
\Lpartband{} frames to the current frames $\overline{\text{PSNR}\left(\sigf_{2t-1},\LP_{2t}\right)}$
for $\reglambda=0$ results from the consideration of the inversion
by our proposed metric $\meshmetricprop$. For the influence of the
regularizer $\regterm$, a value of $\reglambda=0.0004$ yields a
reasonable trade-off between fine-granular mesh deformation, i.e.,
local adaption with respect to the assumptions on the displacement
over time and smoothness $\meshsmoothness$ of the mesh. 

As shown by the example in \fig{\ref{fig:visual-comparison-lowpass}}
on the right, the visual quality is increased significantly by our
proposed metric. There are different reasons for the improvement.
First, by taking the inversion of the compensation into account, another
similarity term is added to the optimization function (\ref{eq:metric13}).
Thereby, the relative influence of the regularizer is decreased. The
impact can be seen by the decrease of the mesh smoothness on the right
half of \tab{\ref{tab:PSNR-in-dB}}. However, this is not the only
reason. Without regularization, our proposed metric is able to increase
the $\overline{\text{PSNR}\left(\sigf_{2t-1},\LP_{2t}\right)}$ from
49.28~dB to 49.70~dB. for the CT volumes. The mesh smoothness also
increases slightly. The inversion of the compensation is also taken
into account by $\meshmetricprop$. Among others, the effect of the
bilinear interpolation is considered which is necessary to obtain
intensity values on the regular pixel as presented in \sect{\ref{sec:Mesh-based-compensation}}.
With our proposed metric~$\meshmetricprop$, the quality of the \Lpartband{}
can
\begin{table*}
\protect\caption{\label{tab:Average-file-size}Average file size per frame in kbyte}

 \setlength\tabcolsep{4pt} 

\begin{center}
\begin{tabular}{c|c|rrr|rrr|rrrr|rrrr}
\toprule
 & & \multicolumn{3}{c|}{$\sigf_t$} & \multicolumn{3}{c|}{no comp}& \multicolumn{4}{c|}{mesh comp, $\meshmetricold$, $\reglambda=0.0004$} & \multicolumn{4}{c}{mesh comp, $\meshmetricprop$, $\reglambda=0.0004$} \tabularnewline
 & & ref $\sigf_{2t-1}$ & cur $\sigf_{2t}$ & sum & $\HP$ & $\LP$ & sum & $\HP$ & $\LP$ & MV & sum & $\HP$ & $\LP$ & MV & sum\tabularnewline

\midrule

\multirow{2}{*}{\begin{turn}{90} CT \end{turn}}
 & \jptwok &   147.36 &   147.36 &   294.72 &   142.06 &        140.96 &        283.02 &        142.00 &  144.06 &  1.98 & 288.04 &  142.67 &  144.80 &  2.36 & 289.83   \tabularnewline

 & \specktwod &    155.56 &   155.56 &   311.12 &   149.52 &        149.58 &        299.09 &        148.98 &  152.19 &  1.98 &  303.15 & 149.40 &  152.75 &  2.36 &  304.51   \tabularnewline


\multirow{2}{*}{\begin{turn}{90} MR \end{turn}}
 & \jptwok &    18.45 &   18.47 &   36.92 &   17.18 &        17.65 &        34.83 &        16.64 &  18.35 &  0.26 &  35.25 &  16.59 &  18.39 &  0.29 &  35.27   \tabularnewline

 & \specktwod &     21.46 &   21.47 &   42.93 &   18.86 &        20.66 &        39.51 &        18.34 &  21.12 &  0.26 &  39.71 & 18.29 &  21.17 &  0.29 &  39.74   

\end{tabular}
\end{center}
\end{table*}
be improved significantly. Furthermore, the similarity to the original
reference frame as well as the current frame can be improved. Using
our proposed metric the average PSNR can be increased from by 47.85~dB
to 48.48~dB by 0.63~dB for the CT volumes.

\begin{figure*}
    \setlength\figureheight{0.32\textwidth}     \setlength\figurewidth{0.32\textwidth}  

\begin{center}

\begin{tabular}{c|c|c}
\multicolumn{2}{c|}{$\reglambda=0.0004$} & $\reglambda=0$\tabularnewline
$\ensuremath{\meshsr}$~0.5 & $\ensuremath{\meshsr}$~0.5, 0.25 & $\ensuremath{\meshsr}$~0.5, 0.25\tabularnewline
$\meshsmoothness=$ \input{img/comparisson/val/vol2_t1-2_z51_m13_LP_hl4_hier_sub1_mesh_smooth_comb.txt} & $\meshsmoothness=$ \input{img/comparisson/val/vol2_t1-2_z51_m13_LP_hl4_hier_sub2_mesh_smooth_comb.txt} & $\meshsmoothness=$ \input{img/comparisson/val/vol2_t1-2_z51_m13_LP_hl4_hier_unc2_mesh_smooth_comb.txt}\tabularnewline
\multicolumn{1}{c}{\input{img/comparisson/vol2_t1-2_z51_m13_mesh_hl4_hier_sub1.tex}} & \multicolumn{1}{c}{\input{img/comparisson/vol2_t1-2_z51_m13_mesh_hl4_hier_sub2.tex}} & \input{img/comparisson/vol2_t1-2_z51_m13_mesh_hl4_hier_unc2.tex}\tabularnewline
\multicolumn{1}{c}{\includegraphics[width=0.31\textwidth]{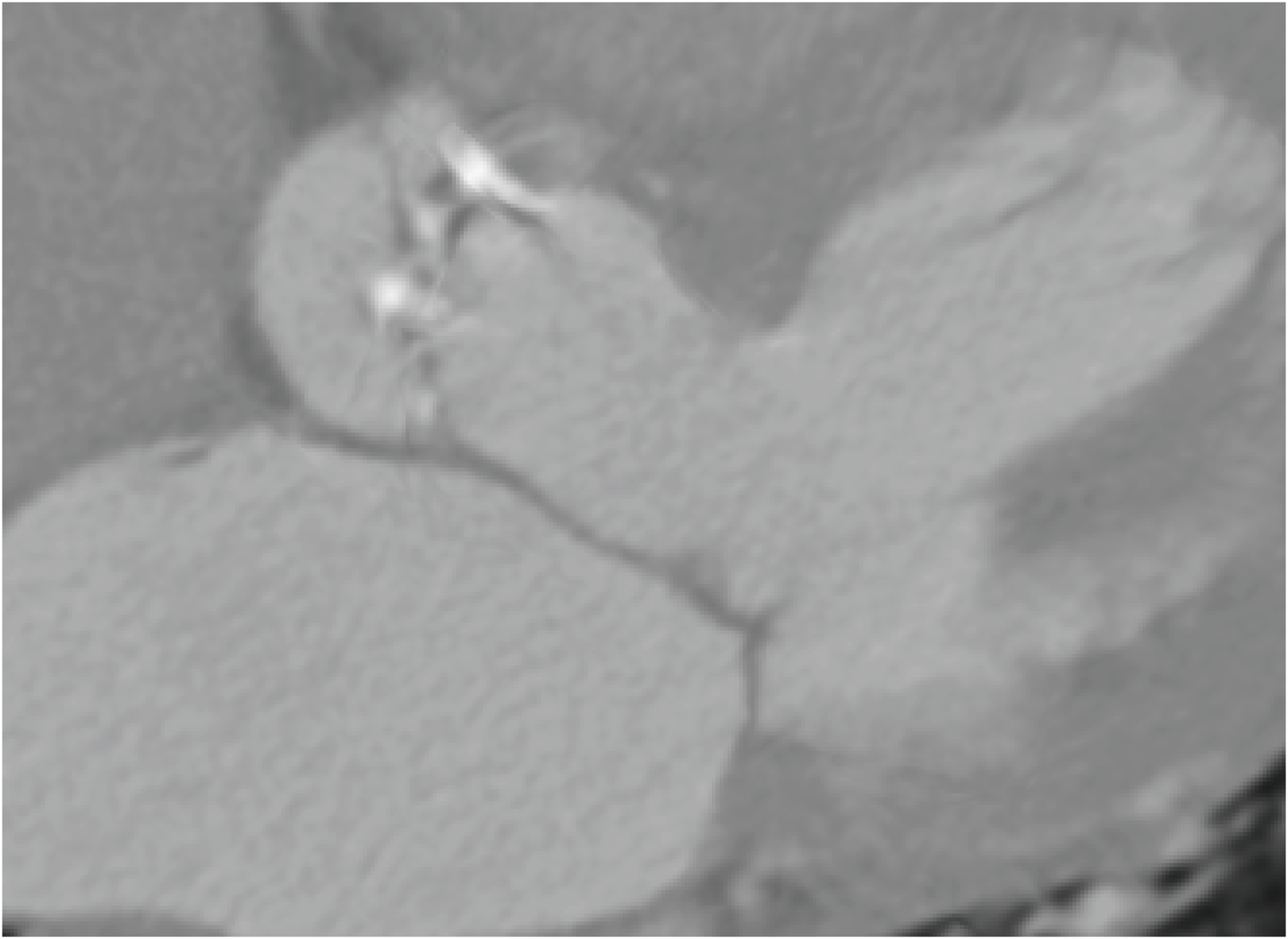}} & \multicolumn{1}{c}{\includegraphics[width=0.31\textwidth]{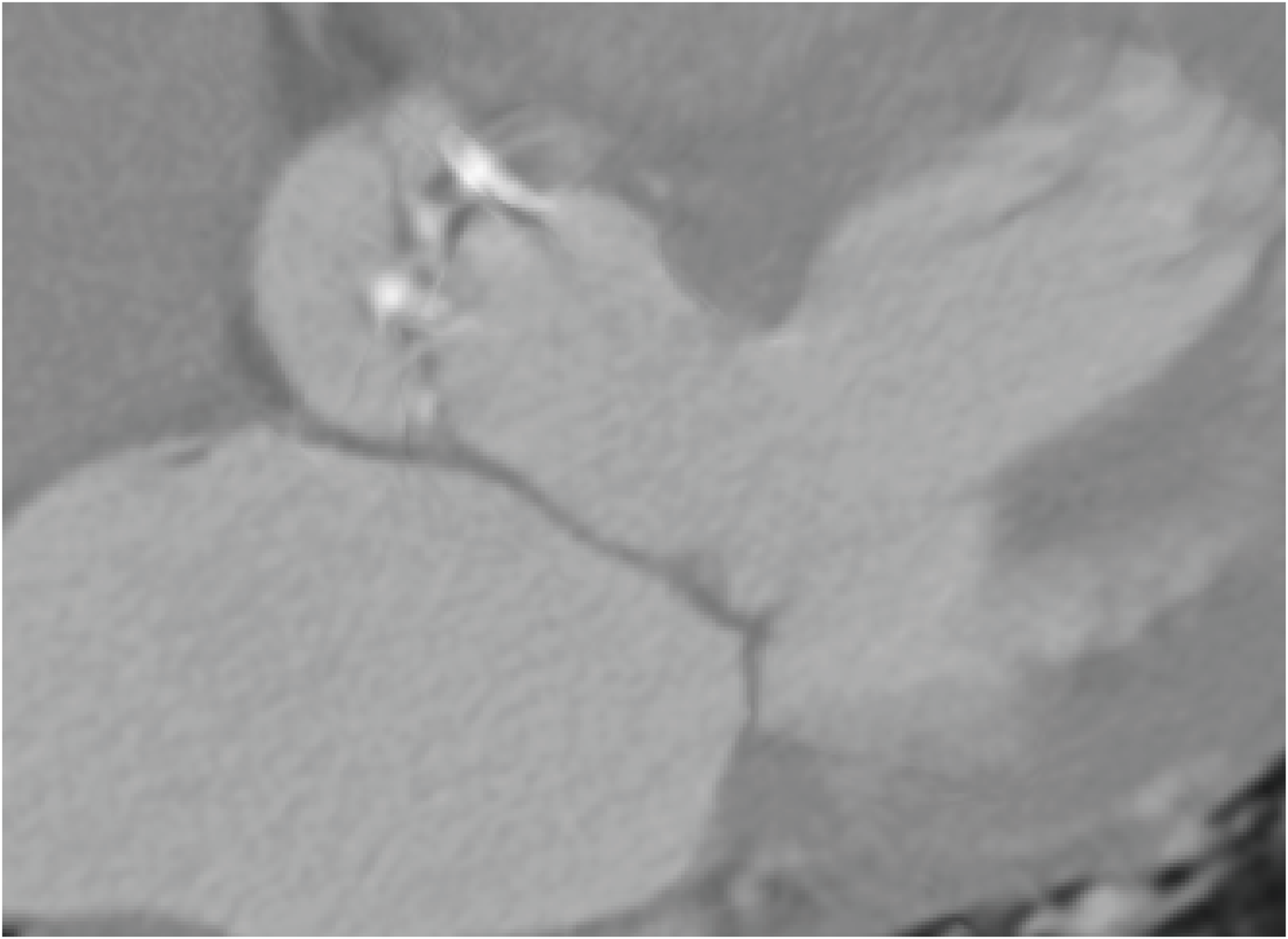}} & \includegraphics[width=0.31\textwidth]{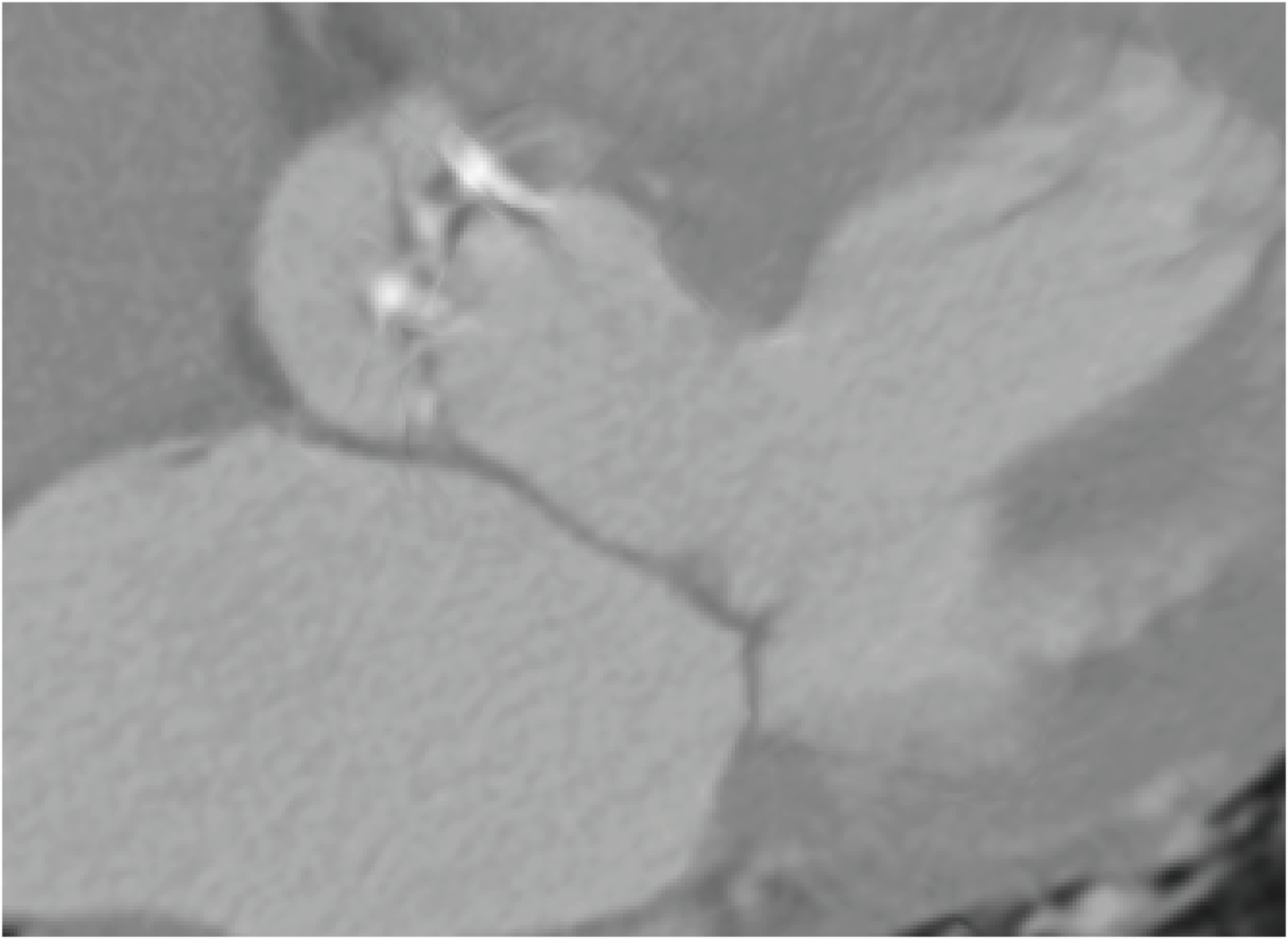}\tabularnewline
PSNR$\left(\sigf_{1},\LP_{1}\right)=$ \input{img/comparisson/val/vol2_t1-2_z51_m13_LP_hl4_hier_sub1_psnr_ref_LP.txt} & PSNR$\left(\sigf_{1},\LP_{1}\right)=$ \input{img/comparisson/val/vol2_t1-2_z51_m13_LP_hl4_hier_sub2_psnr_ref_LP.txt} & PSNR$\left(\sigf_{1},\LP_{1}\right)=$ \input{img/comparisson/val/vol2_t1-2_z51_m13_LP_hl4_hier_unc2_psnr_ref_LP.txt}\tabularnewline
PSNR$\left(\sigf_{2},\mathcal{W}_{1\rightarrow2}\left(\LP_{1}\right)\right)=$
\input{img/comparisson/val/vol2_t1-2_z51_m13_LP_hl4_hier_sub1_psnr_cur_LPMC.txt} & PSNR$\left(\sigf_{2},\mathcal{W}_{1\rightarrow2}\left(\LP_{1}\right)\right)=$
\input{img/comparisson/val/vol2_t1-2_z51_m13_LP_hl4_hier_sub2_psnr_cur_LPMC.txt} & PSNR$\left(\sigf_{2},\mathcal{W}_{1\rightarrow2}\left(\LP_{1}\right)\right)=$
\input{img/comparisson/val/vol2_t1-2_z51_m13_LP_hl4_hier_unc2_psnr_cur_LPMC.txt}\tabularnewline
\end{tabular}

\end{center}

\protect\caption{\label{fig:Visual-extension}Visual comparison for computing $\LP_{1}$,
$z=51$ of the cardiac CT volume using the subpixel estimation in
addition to the hierarchic estimation.\protect \\
The first row shows the deformed meshes resulting from the estimation
using a subpixel search range in addition to the hierarchic estimation.
The images in the second row show details form the corresponding \Lpartbands{}
$\LP_{1}$. On the right, the regularizer has been disabled ($\reglambda=0$)
for these final two refinement iterations. For all cases, the proposed
metric $\meshmetricprop$ has been used.}
\end{figure*}

\begin{table}
\protect\caption{\label{tab:ext-psnr}PSNR in dB and mesh smoothness $\meshsmoothness$
for the subpixel extensions. For comparison, the first column lists
the results for the hierarchic estimation. The next two columns list
the results for the subpixel extension with regularizer enabled. The
last two columns list the results for disabling the regularizer for
the final two iterations.}

 \setlength\tabcolsep{4pt} 

\begin{center}
\begin{tabular}{c|c|r|rr|rr}
\toprule

 & & \multicolumn{5}{c}{proposed metrik $\meshmetricprop$} \tabularnewline
 & & \multicolumn{3}{c|}{$\reglambda=0.0004$} & \multicolumn{2}{c}{$\reglambda=0$} \tabularnewline

 & & hier.  & \multicolumn{2}{c|}{extension} & \multicolumn{2}{c}{extension} \tabularnewline

 & & est.  &
$\meshsr$ 0.5 &
$\meshsr$ 0.5, &
$\meshsr$ 0.5 &
$\meshsr$ 0.5, \tabularnewline
 & & & & 0.25 & & 0.25 \tabularnewline

\midrule
\multirow{3}{*}{\begin{turn}{90} CT \end{turn}}
 & $\overline{\text{PSNR}\left(\sigf_{2t-1},\LP_{t}\right)}$ & 48.48 & 48.60 & 48.66 & 48.94 & 49.12  \tabularnewline

 & $\ensuremath{\overline{\text{PSNR}\hspace{-1mm}\left(\hspace{-0.5mm}\sigf_{2t},\hspace{-0.5mm}\mathcal{W}_{2t\hspace{-0.5mm}-\hspace{-0.5mm}1\hspace{-0.5mm}\rightarrow\hspace{-0.5mm}2t}\hspace{-1mm}\left(\LP_{t}\hspace{-0.5mm}\right)\hspace{-0.5mm}\right)}}$ & 47.60 & 47.70 & 47.75 & 47.95 & 48.10  \tabularnewline

 & $\meshsmoothness$ & 0.820 & 0.822 & 0.828 & 0.711 & 0.665  \tabularnewline

\multirow{3}{*}{\begin{turn}{90} MRT \end{turn}}

 & $\overline{\text{PSNR}\left(\sigf_{2t-1},\LP_{t}\right)}$ & 49.13 & 49.39 & 49.55 & 50.20 & 50.66  \tabularnewline

 & $\ensuremath{\overline{\text{PSNR}\hspace{-1mm}\left(\hspace{-0.5mm}\sigf_{2t},\hspace{-0.5mm}\mathcal{W}_{2t\hspace{-0.5mm}-\hspace{-0.5mm}1\hspace{-0.5mm}\rightarrow\hspace{-0.5mm}2t}\hspace{-1mm}\left(\LP_{t}\hspace{-0.5mm}\right)\hspace{-0.5mm}\right)}}$ & 48.32 & 48.48 & 48.55 & 48.12 & 48.13  \tabularnewline

 & $\meshsmoothness$ & 0.897 & 0.898 & 0.903 & 0.818 & 0.799 

\end{tabular}
\end{center}
\end{table}

\begin{table*}
\protect\caption{\label{tab:ext-filesize}Average file size per frame in kbyte for
the two subpixel extensions}

 \setlength\tabcolsep{4pt} \begin{center}
\begin{tabular}{c|c|rrrr|rrrr|rrrr|rrrr}
\toprule
 & & \multicolumn{8}{c|}{$\reglambda=0.0004$} & \multicolumn{8}{c}{$\reglambda=0$} \tabularnewline
 & & \multicolumn{4}{c|}{subpixel 0.5} & \multicolumn{4}{c|}{subpixel 0.5 0.25}& \multicolumn{4}{c|}{subpixel 0.5} & \multicolumn{4}{c}{subpixel 0.5 0.25} \tabularnewline
 & & $\HP$ & $\LP$ & MV & sum & $\HP$ & $\LP$ & MV & sum & $\HP$ & $\LP$ & MV & sum & $\HP$ & $\LP$ & MV & sum\tabularnewline

\midrule

\multirow{2}{*}{\begin{turn}{90} CT \end{turn}}
 & \jptwok &    141.73 &   144.10 &   3.48 &   289.31 &   142.11 &        144.92 &        4.84 &        291.87 &        142.15 &  144.99 &  4.67 &  291.81 & 143.59 &  146.03 &  6.72 &  296.34   \tabularnewline

 & \specktwod &    148.73 &   152.23 &   3.48 &   304.43 &   148.85 &        152.88 &        4.84 &        306.56 &        148.79 &  152.98 &  4.67 &  306.44 & 149.81 &  153.76 &  6.72 &  310.29   \tabularnewline


\multirow{2}{*}{\begin{turn}{90} MR \end{turn}}
 & \jptwok &     16.58 &   18.29 &   0.46 &   35.33 &   16.49 &        18.29 &        0.58 &        35.36 &        16.48 &  18.45 &  0.58 &  35.51 & 16.45 &  18.42 &  0.77 &  35.64   \tabularnewline

 & \specktwod &     18.28 &   21.14 &   0.46 &   39.88 &   18.18 &        21.21 &        0.58 &        39.96 &        18.17 &  21.30 &  0.58 &  40.04 & 18.12 &  21.33 &  0.77 &  40.22   

\end{tabular}
\end{center}
\end{table*}

To evaluate the compressibility of the resulting sub-bands from the
compensated \wt{} in temporal direction, the sub-bands have been
coded lossless using two standard state of the art wavelet coefficient
coders, namely JPEG2000 \cite{christopoulos2002} and SPECK \cite{SPECKPatent}.
Therefore, an additional 5~level dyadic decomposition in $xy$-direction
of each frame was applied using the \selectlanguage{english}%
\Legall{}\selectlanguage{american}%
. \Tab{\ref{tab:Average-file-size}} lists the resulting average file
size per frame in kbyte for the \Hpartband{} and the \Lpartband{}
to see the contribution of the sub-bands to the overall file size.
The rate is obtained using \cite{fowler2000qccpack}. Overall, both
methods perform in a very similar way. A wavelet decomposition in
temporal direction leads to better results than coding a volume slice-by-slice.
The best coding results can be obtained by applying a \wt{} in temporal
direction without compensation. However, uncompensated displacement
causes artifacts in the \Lpartband{} as shown in \fig{\ref{fig:Visual-orig-haar}}.
Lossless reconstruction is a crucial condition for medical image data.
Due to the data acquisition process, noisy structures are very similar
in neighboring frames. Usually a transform is applied along neighboring
pixels. By incorporating a compensation method into the transform,
it is applied on the motion trajectory \cite{Ohm1994}. This makes
it difficult to exploit the similar noisy structures for coding \cite{schnurrer2012mmsp},
\cite{schnurrer2014}. The results for the sub-bands from the compensated
transform are in a similar range. For obtaining a high quality \Lpartband{},
the rate increases by about 2.4\% compared to uncompensated \wt{}.

In coincidence with the subpixel estimation for block-based motion
compensation \cite{hevc_overview}, we tested two extensions using
a subpixel $\meshsr$ in the refinement. Therefore, one iteration
after the hierarchic estimation is computed using an $\meshsr=\frac{1}{2}$
followed by an iteration using $\meshsr=\frac{1}{4}$. In the first
extension, the parameter for the regularizer was not changed. In the
second extension the regularizer was deactivated for the final two
iterations. The resulting meshes are shown in \fig{\ref{fig:Visual-extension}}.
In the first column, the first iteration with $\meshsr=\frac{1}{2}$
was computed. In the last two columns, both iterations were computed.
Only the subpixel refinement in the end is considered, therefore the
maximum movement of a grid point equals a distance of 0.75~pixel.
As a result, the meshes look very similar. Nevertheless, performing
the subpixel refinement without regularizer, i.e., $\reglambda=0$,
the resulting mesh on the right is not as smooth as the mesh in the
center. \Tab{\ref{tab:ext-psnr}} lists the results for all simulated
volumes. For comparison, the first column repeats the results from
the hierarchic estimation before the subpixel extensions. The subpixel
extensions can further improve the similarity of the \Lpartband{}
to the original sequence. The smoothness can slightly be improved
if the regularizer is kept enabled. The metric for the smoothness
confirms that disabling the regularizer for the last two iterations
leads to very fine granular deviations in the mesh. This also increases
the file size for lossless coding, listed in \Tab{\ref{tab:ext-filesize}}.
With the regularizer, the file size for the CT volume is reduced slightly
after the first iteration of the subpixel extensions.

Keeping the regularizer enabled for a subpixel refinement as final
step of the iterative estimation process is advantageous for the mesh
smoothness, the compressibility of the sub-bands and the quality of
the \Lpartband{}.

\section{\label{sec:Conclusion}Summary and Conclusion}

A \scalable{} representation of a dynamic volume entails several
advantages. We propose a method to obtain a temporal \scalable{}
representation of dynamic volumes. By incorporating feasible compensation
methods into the \wt{}, the quality of the \Lpartband{} can be improved
by 0.63~dB and 0.43~dB for CT and MR volumes at the cost of a small
rate increase. A high quality of the \Lpartband{} of a compensated
\wt{} is necessary when it shall be utilized as \downscale{}d version
of the signal.

In addition to the compensation in the prediction step, the inversion
in the update step is also a crucial part of compensated \wl{}. Current
research has not considered the inversion of the compensation during
the estimation. We added this missing part to the model for finding
the optimum parameters for the compensation method. The \Lpartband{}
can be additionally improved by extending the metric, to serve as
high quality \scalable{} representation with smaller temporal resolution.

\section*{Acknowledgment}

The authors gratefully acknowledge that this work has been supported
by the Deutsche Forschungsgemeinschaft (DFG) under contract number
KA~926/4-2.

\bibliographystyle{IEEEtran}
\bibliography{bib/bibliography}

\end{document}